\documentclass[twocolumn,prd,aps,nofootinbib,showpacs]{revtex4}

\usepackage{epsf,amsmath,amssymb,mathptmx}

\newcommand{\FIGURE}[2][v]{\begin{figure}[#1]#2
                                        \end{figure}}
\newcommand{\FIGURESTAR}[2][v]{\begin{figure*}[#1]#2
                                        \end{figure*}}
\newcommand{\vfs}{{\abbrev VFS}}
\newcommand{\ffs}{{\abbrev FFS}}

\newcommand{\abbrev}{\small}
\newcommand{\ep}{\epsilon}

\newcommand{\api}{\frac{\alpha_s}{\pi}}

\newcommand{\eqn}[1]{Eq.\,(\ref{#1})}
\newcommand{\fig}[1]{Fig.\,\ref{#1}}
\newcommand{\figs}[1]{Figs.\,\ref{#1}}

\newcommand{\sct}[1]{Sect.\,\ref{#1}}
\newcommand{\reference}[1]{Ref.\,\cite{#1}}
\newcommand{\refs}[1]{Refs.\,\cite{#1}}
\newcommand{\dd}{{\rm d}}

\newcommand{\order}[1]{{\cal O}(#1)}

\newcommand{\Lx}{\left(}
\newcommand{\Rx}{\right)}
\newcommand{\LB}{\left[}
\newcommand{\RB}{\right]}
\newcommand{\Li}[1]{{\mathop{\rm Li}_{#1}\nolimits}}
\newcommand{\Di}[1]{{\cal D}_{#1}}

\newcommand{\lo}{{\abbrev LO}}
\newcommand{\nlo}{{\abbrev NLO}}
\newcommand{\nnlo}{{\abbrev NNLO}}

\newcommand{\lnbm}{l_{b}}

\newcommand{\muR}{\mu_R}
\newcommand{\muF}{\mu_F}

\newcommand{\msbar}{\overline{\mbox{\small MS}}}

\newcommand{\higgs}{\phi}

\newcommand{\mhiggs}{M_\higgs}

\newcommand{\dglap}{{\abbrev DGLAP}}
\newcommand{\mssm}{{\rm\abbrev MSSM}}
\newcommand{\qcd}{{\abbrev QCD}}
\newcommand{\rge}{{\abbrev RGE}}
\newcommand{\lep}{{\abbrev LEP}}
\newcommand{\lhc}{{\abbrev LHC}}

\newcommand{\bare}{{\rm B}}
\newcommand{\bbar}{b\bar b}
\newcommand{\qqbar}{q\bar q}
\newcommand{\feynsl}[1]{
  \setbox0=\hbox{/} \setbox1=\hbox{$#1$}
  \dimen0=\wd0 \advance\dimen0 by -\wd1 \divide\dimen0 by 2
  \ifdim\wd0>\wd1 \lower.15ex
          \copy0\kern-\wd0\kern\dimen0\copy1\kern\dimen0
  \else \kern-\dimen0\lower.15ex
          \copy0\kern-\dimen0\kern-\wd1\copy1\fi}
\begin{document}
\pacs{14.80.Bn, 14.80.Cp, 12.38.-t, 12.38.Bx}
\begin{titlepage}
\hspace*{\fill}\parbox[t]{4cm}{
BNL-HET-03/4\\
CERN-TH/2003-067}

\bibliographystyle{JHEP}

\title{Higgs boson production in bottom quark fusion at
  next-to-next-to-leading order}

\author{Robert V. Harlander}
\email[Email:]{robert.harlander@cern.ch}
\affiliation{TH Division, CERN, CH-1211 Geneva, Switzerland}
\author{William B. Kilgore}
\email[Email:]{kilgore@bnl.gov}
\affiliation{Physics Department, Brookhaven National Laboratory,
      Upton, NY 11973, U.S.A.}
\date{1 July 2003}
\preprint{BNL-HET-03/4}
\preprint{CERN-TH/2003-067}
\eprint[]{hep-ph/0304035}

\begin{abstract}
  The total cross section for Higgs production in bottom-quark
  annihilation is evaluated at next-to-next-to-leading order
  in \qcd{}. This is the first time that all terms at order
  $\alpha_s^2$ are taken into account. We find a greatly reduced scale
  dependence with respect to lower order results, for both the
  factorization and the renormalization scales. The behavior of the
  result is consistent with earlier determinations of the appropriate
  factorization scale for this process of $\muF\approx M_H/4$, and
  supports the validity of the bottom parton density approach for
  computing the total inclusive rate.  We present precise predictions
  for the cross section at the Fermilab Tevatron and the CERN Large
  Hadron Collider.
\end{abstract}

\maketitle
\end{titlepage}

\section{Introduction}

The search for the Higgs boson will be a top priority of the CERN
Large Hadron Collider (\lhc{}).  The \lhc's discovery potential for
the Standard Model Higgs boson fully covers the mass range from the
experimental lower bound established by the \lep{} experiments
($M_H\gtrsim 114$\,GeV) up to $M_H\approx 1$\,TeV, beyond which the
concept of the Higgs boson as an elementary particle becomes
questionable.  In addition, the Fermilab Tevatron could find evidence
for or even discover the Higgs boson if $M_H\lesssim 180$\,GeV and if
sufficient luminosity~\cite{lumimoni} can be collected.

The theoretical description of the signal processes for Standard Model
Higgs boson production is under good control. For a review,
see~\reference{Carena:2002es}. The dominant production mode is gluon
fusion, for which the next-to-next-to-leading order (\nnlo{})
corrections are now available
\cite{Harlander:2002wh,Anastasiou:2002yz} and have recently been
reconfirmed by \reference{Ravindran:2003um}.  The radiative
corrections for the weak boson fusion channel~\cite{Han:1992hr} and
associated production with a weak gauge boson~\cite{Han:1991ia} have
been known for several years, rendering the theoretical uncertainty in
these processes very small.  Recently, next-to-leading order (\nlo{})
corrections have also been evaluated for Higgs boson production in
association with top quarks~\cite{
Reina:2001sf,Beenakker:2001rj,Dawson:2002tg,Beenakker:2002nc},
resulting in a drastic reduction of the scale uncertainty.

These results can be used for supersymmetric Higgs boson production as
well.  However, because of the enriched particle spectrum in
supersymmetric extensions of the Standard Model, they provide only a
part of the full production rate in general.  Additional contributions
arise through intermediate supersymmetric
partners~\cite{Dawson:1996xz} and modified couplings of the Standard
Model particles.  In order to avoid unnecessary generalizations, we
will focus on the Minimal Supersymmetric Standard Model (\mssm{}) for
the rest of this paper (see, e.g., \reference{Martin:1997ns} for an
outline of the \mssm{}).  The extent to which our results can be
transferred to other models should be clear from this discussion.

The \mssm{} contains two Higgs doublets, one giving mass to up-type
quarks and the other to down-type quarks. The associated vacuum
expectation values are labeled $v_u$ and $v_d$, respectively, and
they determine the \mssm{} parameter $\tan\beta\equiv v_u/v_d$.  After
spontaneous symmetry breaking, there are five physical Higgs bosons,
whose mass eigenstates are denoted by $h$ (``light scalar''), $H$
(``heavy scalar''), $H^\pm$ (``charged scalars''), and $A$
(``pseudoscalar''). One interesting consequence of this more
complicated Higgs sector is that, compared to the Standard Model, the
bottom quark Yukawa coupling can be enhanced with respect to the top
quark Yukawa coupling.  In the Standard Model, the ratio of the $t\bar
tH $ and $\bbar H$ couplings is given at the tree-level by
$\lambda^{\rm SM}_{t}/\lambda^{\rm SM}_{b} = m_t/m_b\approx 35$. In
contrast, in the \mssm{}, it depends on the value of $\tan\beta$.  At
leading order,
\begin{equation}
\label{eq::Yukrat}
\frac{\lambda^{\mssm}_{t}}{\lambda^{\mssm}_{b}} =
f_\higgs(\alpha)\,\frac{1}{\tan\beta}\cdot \frac{m_t}{m_b}\,,
\end{equation}
with
\[
f_\higgs(\alpha) = \left\{
  \begin{array}{ll}
    -\cot\alpha\,,& \qquad \higgs=h\,,\\
    \phantom{-}\tan\alpha\,,& \qquad \higgs=H\,,\\
    \phantom{-}\cot\beta\,, & \qquad \higgs=A\,,
  \end{array}
  \right.
\]
\FIGURE[t]{
    \begin{tabular}{cc}
    \epsfxsize=3.6cm
    \epsffile[145 450 460 660]{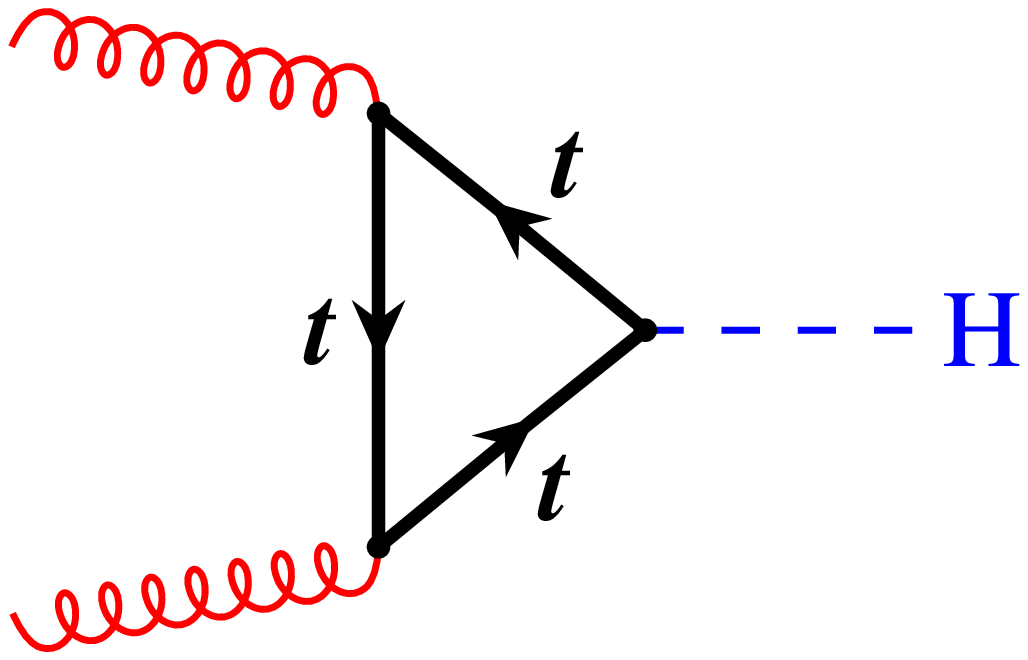}&
    \epsfxsize=3.6cm
    \epsffile[145 450 460 660]{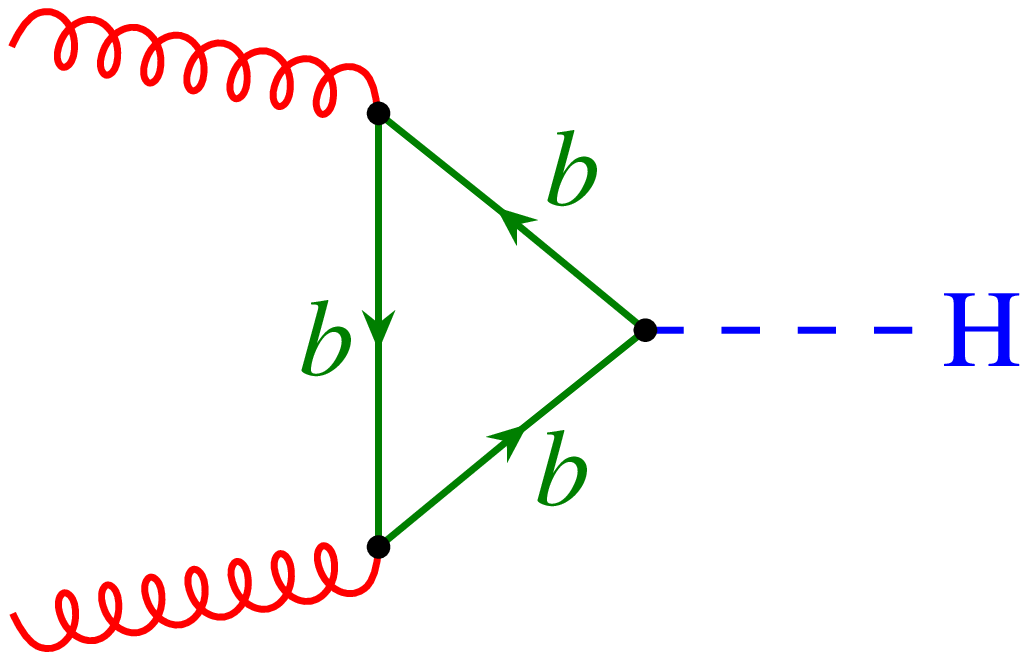}\\
   (a) & (b)
    \end{tabular}
    \caption[]{\label{fig::bloop}For large $\tan\beta$, the bottom quark
      contribution to
      the gluon fusion process can be comparable to the top quark
      contribution.}  }
where $\alpha$ is the mixing angle between the weak and the mass
eigenstates of the neutral scalars.  A value of $\tan\beta$ as large
as $30-40$ could be accommodated fairly naturally in the \mssm{}.
Such an enhancement would have (at least) two important consequences.
The first is that in the gluon fusion mode it is no longer sufficient
to consider top quark loops as the only mediators between the Higgs
boson and the gluons; one must also include the effects of bottom
quark loops (see \fig{fig::bloop}).  Since one cannot justify the use
of an effective field theory in which the bottom quarks are integrated
out, this involves computing massive multi-loop diagrams.  While the
massive \nlo{} calculation (including massive two-loop virtual
diagrams) was performed some time ago~\cite{Spira:1995rr}, the \nnlo{}
result (requiring up to three loops for the virtual correction) is
still beyond the limits of current calculational technology.

The second consequence is that Higgs boson production in association
with bottom quarks can become an important channel: $p\tilde{p}
\to\bbar\higgs$ ($\tilde{p}\in\{p,\bar p\}$ and $\phi\in\{h,H,A\}$
here and in what follows).  At first sight, the evaluation of the
corresponding cross section is in close analogy to the process
$p\tilde{p}\to t\bar t\phi$. But this is only true if the bottom quarks
are observed in the detector, and are thus restricted to large
transverse momenta. If at least one of the bottom quarks escapes
detection, the production rate must be integrated over all transverse
momenta of this bottom quark. Since the Higgs boson is much heavier
than the bottom quark, this integration leads to collinear logarithms,
$\ln(m_b/\mhiggs)$, which require a more careful analysis than in
the case of $t\bar t\higgs$ production.

The subject of this paper is the precise prediction of the total cross
section for Higgs boson production in association with bottom quarks,
where neither bottom quark need be detected.  This requires
integrating over the transverse momenta of {\it both} final state
bottom quarks.  Each integration gives rise to collinear logarithms of
the kind mentioned above. Since the bottom quarks may remain
undetected, it is more appropriate to view our result as a part of the
total inclusive Higgs production rate $\sigma(p\tilde{p}\to
\higgs+X)$. In order to emphasize this point, we shall henceforth
denote the fully inclusive process mediated through
bottom--antibottom annihilation as $p\tilde{p}\to (\bbar)\higgs+X$.

Our calculation is based on the approach of
Refs.\,\cite{Dicus:1989cx,Dicus:1998hs,Maltoni:2003pn}, where the
leading-order (\lo{}) partonic process is taken to be $\bbar\to
\higgs$.  We will refer to this as the {\it variable flavor number
scheme\/} (\vfs{}) approach in what follows, as opposed to the {\it
fixed flavor number scheme\/} (\ffs{}) approach, where the tree-level
process $gg\to \bbar\higgs$ is taken as the lowest-order contribution
and bottom quarks cannot appear in the initial state. The initial
state bottom quarks in the \vfs{} approach arise (predominantly) from
gluon splitting in the proton, parametrized in terms of bottom quark
parton distributions~\cite{Barnett:1988jw,Olness:1988ep,
Aivazis:1994pi,Thorne:1998ga,Kramer:2000hn,Buza:1996ie,Buza:1998wv,
Chuvakin:1999nx,Chuvakin:2000qc,Chuvakin:2000jm,Chuvakin:2000zj}.
In this way, the large collinear logarithms that arise due to the fact
that the colliding gluons carry a momentum of the order of
$\mhiggs/2\gg m_b$ can be resummed through {\dglap} evolution.  The
convolution of these bottom quark densities with the partonic cross
section leads to the hadronic cross section $\sigma\LB p\tilde{p} \to
(\bbar)\higgs+X\RB$.

This process has been a subject of interest for some time. It is
currently known up to \nlo{} in the \vfs{} approach
\cite{Eichten:1984eu,Gunion:1987pe,Olness:1988ep,Dicus:1989cx,
Dicus:1998hs,Maltoni:2003pn}.  In the \ffs{} approach, the calculation
is analogous to $t\bar t\higgs$ production, which is also known to
\nlo{}~\cite{Raitio:1979pt,Ng:1984jm,Kunszt:1984ri,Reina:2001sf,
Beenakker:2001rj,Dawson:2002tg,Beenakker:2002nc}.  The case where one
bottom quark is tagged has been computed at \nlo{} in
\reference{Campbell:2002zm}; in that case, the \lo{} process in the
\vfs{} approach is $bg\to b\higgs$.

There has been an ongoing discussion as to the relative merits of the
\vfs{} and \ffs{}
approaches~\cite{Dicus:1998hs,Balazs:1998sb,Campbell:2002zm,
Maltoni:2003pn,Rainwater:2002hm,Spira:2002rd}, especially because the
results of the two approaches disagree by more than an order of
magnitude for the scale choice $\muF=\muR=\mhiggs$, where $\muF$ and
$\muR$ are the factorization and the renormalization scales,
respectively.  Recently, it has been argued~\cite{Maltoni:2003pn},
that the proper factorization scale for this process should be
$\muF\approx\mhiggs/4$ instead of $\mhiggs$. Indeed, for this choice,
the disagreement between the \vfs{} and the \ffs{} approach is
significantly reduced.  The result of our paper demonstrates the
self-consistency of the \vfs{} approach and confirms the proposed
factorization scale of \reference{Maltoni:2003pn} as the appropriate
choice.

Given the considerations above, the motivations for a \nnlo{}
calculation are manyfold.  One is to examine the assertion that
$\muF\approx\mhiggs/4$ is the proper choice for this process at
\nlo~\cite{Plehn:2002vy,Maltoni:2003pn}.  If the higher order
corrections are minimal at that scale, this would be a strong argument
in favor of the validity of the \vfs{} approach to $(\bbar)\higgs$
production.  A second motivation, as will be discussed in more detail
below, is that the \nnlo{} terms play an exceptional role in the
\vfs{} approach to $(\bbar)\higgs$ production due to the fact that
they are the first to consistently include the ``parent'' process,
$gg\to\bbar\higgs$, and thereby sample the same range of bottom quark
transverse momenta as the \lo{} \ffs{} approach.  A third and perhaps
dominant motivation for the \nnlo{} calculation is to reduce the
sensitivity of the calculation to the unphysical scale parameters
$\muF$ and $\muR$, thereby removing a significant source of
uncertainty from the theoretical prediction.

In this paper we will present results for the process $p\tilde{p}\to
(\bbar)\higgs+X$ at \nnlo{}. As will be shown, they nicely meet all
expectations concerning their dependence on the renormalization and
factorization scales, thus providing a solid prediction for the total
cross section at the \lhc{} and the Tevatron.  The inclusive
production cross section could have phenomenological implications for
the observation of the supersymmetric $H$ and $A$ bosons, for example,
in the $H/A\to\mu^+\mu^-$ decay mode.

The organization of the paper is as follows:
In Sec.~\ref{sec:bdist} we discuss the \vfs{} approach to computing the
$p\tilde{p}\to (\bbar)\higgs+X$ process and its motivations.  In
Sec.~\ref{sec:outline} we describe the actual calculation and in
Sec.~\ref{sec:results} we present our numerical results.  Analytic
results for the partonic cross sections are presented in the appendix.



\FIGURE[ht]{
\begin{tabular}{cc}
\kern -25pt
\epsfxsize=3.6cm
\epsffile[150 500 410 670]{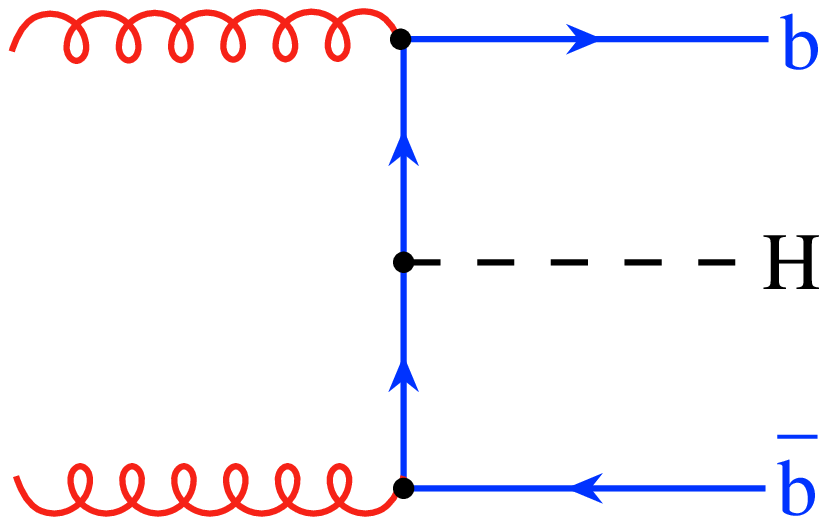} &
\epsfxsize=3.96cm
\epsffile[115 520 380 680]{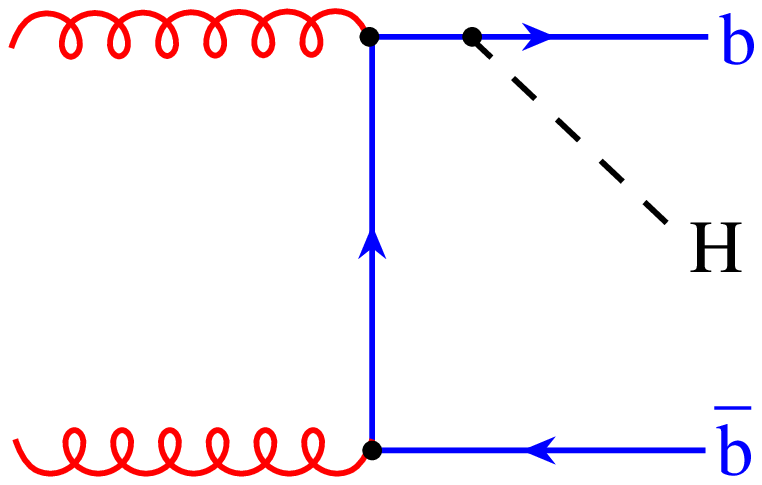} \\
(a) & (b) \\[2em]
\kern -25pt
\epsfxsize=3.96cm
\epsffile[115 520 380 680]{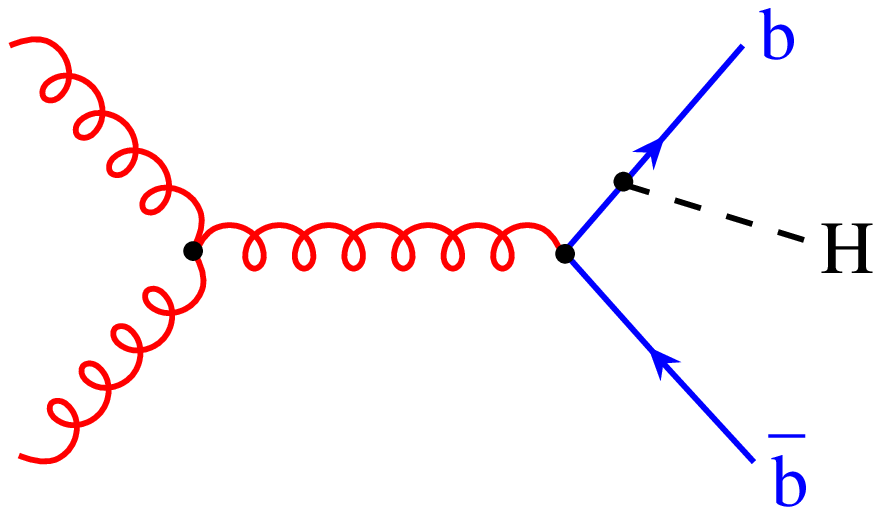} &
\epsfxsize=3.96cm
\epsffile[115 520 380 680]{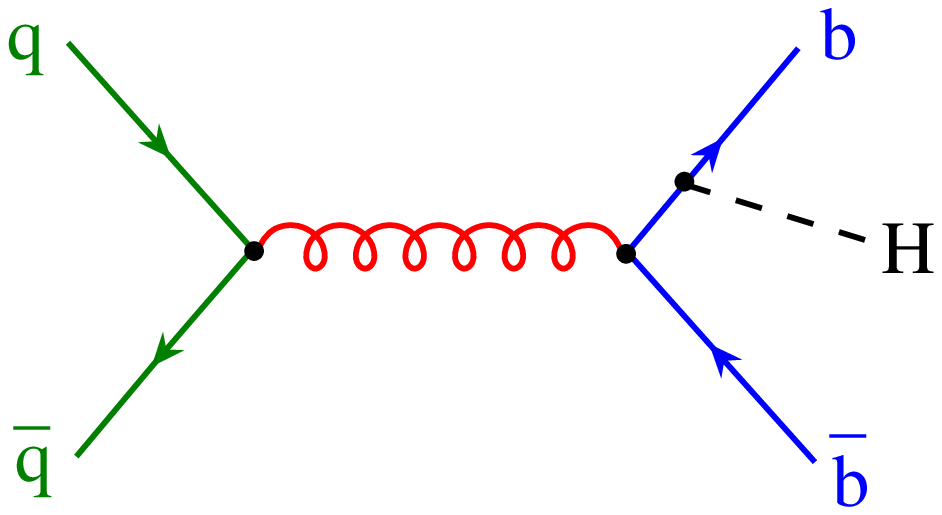} \\
(c) & (d) \\[2em]
\end{tabular}
\caption[]{\label{fig::gghbb} Partonic processes for $pp\to\bbar H$.
  Not shown are diagrams that can be obtained by crossing the initial
  state gluons, or radiating the Higgs off an antibottom quark.}
}

\section{Theoretical description of the production rate}
\label{sec:bdist}
In the \ffs{} approach, the \lo{} partonic process for the production
of a Higgs boson in association with a bottom quark pair is of order
$\alpha_s^2$. A few typical diagrams are shown in \fig{fig::gghbb}.
Because of the large mass difference between the bottom quark and the
Higgs boson, the total cross section contains large logarithms of the
form
\begin{equation}
\begin{split}
\lnbm \equiv \ln(m_b^2/\mu_\phi^2)\,,
\end{split}
\end{equation}
where $\mu_\phi$ is of the order of $\mhiggs$.  More precisely, every
on-shell gluon that splits into a $\bbar$ pair with an on-shell bottom
quark generates one power of that logarithm.  Thus,
\figs{fig::gghbb}(a) and \ref{fig::gghbb}(b) generate two and one
power of $\lnbm$, respectively, while \figs{fig::gghbb}(c) and
\ref{fig::gghbb}(d) do not generate any $\lnbm$ terms.  Furthermore,
each higher order in perturbation theory brings in another power of
$\lnbm$ due to the radiation of gluons from bottom quarks.

Because $\lnbm\sim \ln(m_b^2/\mhiggs^2)$ is rather large,
$\alpha_s\lnbm$ is not a good expansion parameter.  It would be better
to re-organize the perturbative series such that terms like
$(\alpha_s\lnbm)^n$ are resummed to all orders in $n$.  This
resummation can be achieved by introducing bottom quark parton
distribution functions which contain all the collinear terms arising
from the splitting of gluons into $\bbar$ pairs~\cite{Barnett:1988jw,
Olness:1988ep,Aivazis:1994pi,Thorne:1998ga,Kramer:2000hn,Buza:1996ie,
Buza:1998wv,Chuvakin:1999nx,Chuvakin:2000qc,Chuvakin:2000jm,
Chuvakin:2000zj}.  This constitutes the motivation for using the
\vfs{} approach.

Convolving the tree-level process of \fig{fig::real1}(a) with these
bottom quark distributions will resum the leading logarithms of the
form $(\alpha_s\lnbm)^2\cdot(\alpha_s\lnbm)^{n}$, $n\geq 0$. In order
to retain subleading logarithms, one has to compute higher
orders. For example, including the \nlo{} contributions with all the
relevant subprocesses ($\bbar\to h$, $\bbar\to hg$, $gb\to hb$, and
$g\bar b\to h\bar b$), resums the terms of order
$\alpha_s^2\lnbm\,(\alpha_s\lnbm)^{n}$, $n\geq 0$. In order to
retain {\it all} powers of $\lnbm$ at order
$\alpha_s^2\,\sum_n(\alpha_s\lnbm)^{n}$, it is necessary to evaluate
the cross section up to \nnlo{}.

\FIGURESTAR[ht]{
\begin{tabular}{ccc}
\epsfxsize=4cm
\raisebox{-1em}{\epsffile[150 460 390 670]{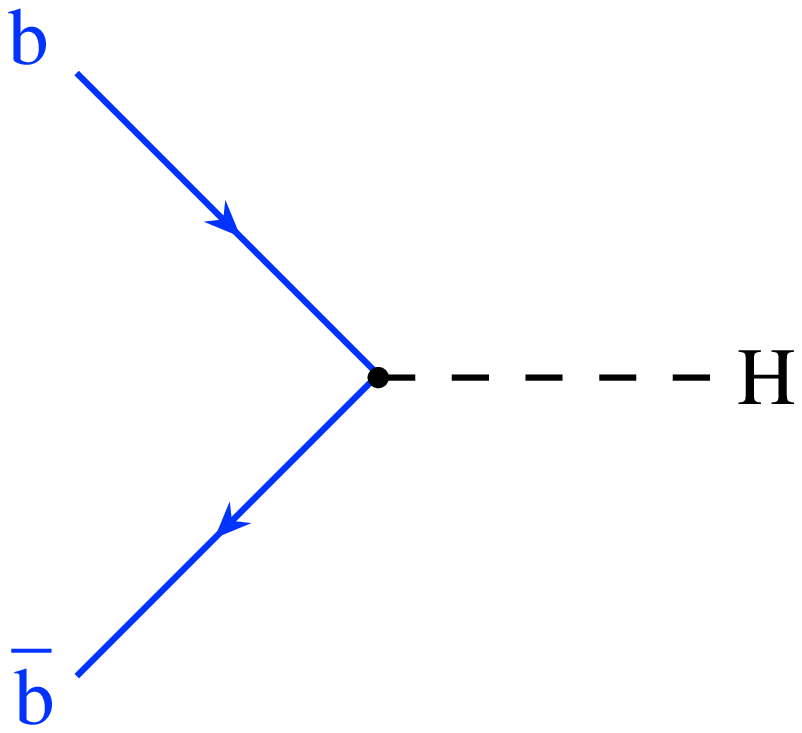}}\ \ &
\epsfxsize=4cm
\epsffile[150 500 410 670]{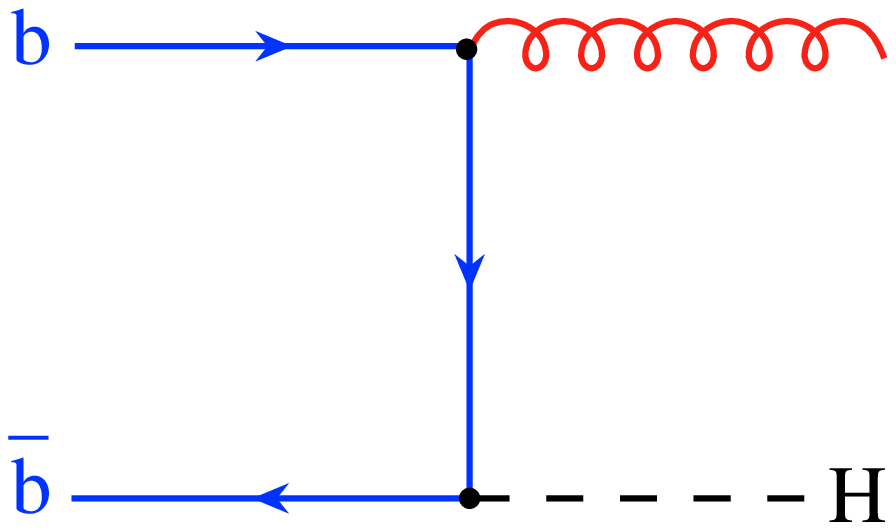}\ \ &
\epsfxsize=4cm
\epsffile[150 500 410 670]{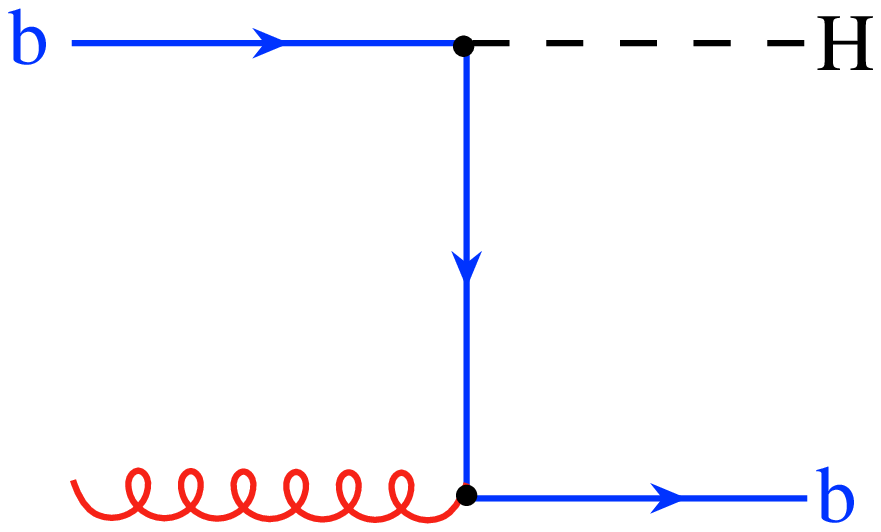}\ \ \\
(a) & (b) & (c)
\end{tabular}
\caption[]{\label{fig::real1}Lowest order diagrams contributing to 
  (a) $b\bar b\to H$, (b) $b\bar b\to Hg$, and (c) $bg\to Hb$. At
  \nnlo{}, these diagrams receive corrections up to two loops in case
  (a), and one loop in case (b) and (c).}
}

Let us briefly review the idea of the \vfs{} in its simplest form.
Assume $n_\ell=n_f-1$ massless quark flavors and one massive quark
flavor of mass $m_h$. First, one defines parton densities
$f^{(n_\ell)}_{i}(x,Q^2)$ for the gluon ($i=g$) and the massless
flavors ($i=1,\ldots,n_\ell$) in the standard way, obeying \dglap{}
evolution with $n_\ell$ active flavors. The heavy quark density
$f^{(n_\ell)}_{n_f}(x,Q^2)$ is assumed to vanish. Partonic processes
involving the heavy quark should be evaluated by keeping the heavy
quark mass. This is called the $n_\ell$-flavor scheme.

At a certain scale $\mu_h^2$, one relates the $n_\ell$- to the
$n_f$-flavor scheme by defining initial conditions for new parton
densities $f^{(n_f)}_i$ in terms of the $f^{(n_\ell)}_i$:
\begin{equation}
\begin{split}
  f^{(n_f)}_{i}(x,Q^2=\mu_h^2) &= \sum_j
        C_{ij}(\mu_h/m_h)\otimes f^{(n_\ell)}_{j}(x,Q^2=\mu_h^2)\,,\\
  i&=g,1,\ldots,n_f \qquad j=g,1,\ldots,n_\ell\,.
\label{eq::fini}
\end{split}
\end{equation}
The $C_{ij}$ are determined by the requirement that physical
quantities are the same (up to higher orders in $\alpha_s$) in both
the $n_\ell$- and the $n_f$-flavor scheme.  (This requirement may be
implemented asymptotically or using mass dependent
terms~\cite{Aivazis:1994pi,Thorne:1998ga,Buza:1998wv,Chuvakin:1999nx}.)
Above the matching scale, the {\dglap} evolution of the
$f^{(n_f)}_i(x,Q^2)$ ($i=g,1,\ldots,n_f$) is performed with $n_f$
active flavors.

In general, one assumes the $n_\ell$-flavor scheme at scales $Q^2\lesssim
m_h^2$ and switches to the $n_f$-flavor scheme at larger values of
$Q^2$. It is also convenient to choose the matching scale in
\eqn{eq::fini} as $\mu_h^2=m_h^2$, which avoids the occurrence of
logarithms of $m_h/\mu_h$.

For our purposes, $m_h^2/Q^2\sim m_b^2/\mhiggs^2\lesssim 0.003$, so
threshold effects from the matching prescription should be minimal.
For the same reason, it is justified to neglect the bottom quark mass
in the partonic process (apart from Yukawa couplings, of course).
Indeed, masses must be neglected for initial state bottom quarks in
order to avoid violation of the Bloch-Nordsieck theorem (and the
consequent failure to fully cancel infrared divergences) at \nnlo{}
and beyond~\cite{Catani:1988xy}.

In order to make the following discussion more transparent, let us write
the fully inclusive $(\bbar)\higgs$ production rate in the \vfs{}
approach schematically as follows:
\begin{equation}
\begin{split}
  \sigma(p\tilde{p}\to& (\bbar)\higgs+X) =\\&
  \sum_{n=0}^\infty\left(\alpha_s\lnbm\right)^n\bigg\{
   \alpha_s^2\,\left[
    c_{n0}\,\lnbm^2 + c_{n1}\,\lnbm + c_{n2}\right]\\
   &\qquad+ \alpha_s^3\,d_{n3}
   + \alpha_s^4\,d_{n4}
   + \alpha_s^5\,d_{n5} +\cdots\bigg\}\,.
   \label{eq::sums}
\end{split}
\end{equation}
The sum over $n$ is implicit in the parton densities.  A \lo{}
calculation determines the coefficients $c_{n0}$, while \nlo{} adds
the coefficients $c_{n1}$. Note that the subprocess $bg\to b\higgs$
does not fully determine the coefficients $c_{n1}$; in order to obtain
the correct resummation at $\alpha_s^n\lnbm^{n-1}$ ($n\geq 2$), one
has to include the real and virtual corrections to the $\bbar\to
\higgs$ subprocess as well.  In the same way, the sum of all
subprocesses that contribute at second order determines $c_{n2}$
($n\geq 0$), and thus {\it all terms associated with the order
$\alpha_s^2\,(\alpha_s\lnbm)^n$}.  The \nnlo{} result is thus the
first to include all terms of order $\alpha_s^2$ (as well as higher
order terms resummed in the parton distribution functions).  Higher
orders in perturbation theory correspond to the coefficients $d_{nk}$;
their $\lnbm$ terms are --- formally --- completely contained in the
parton densities.  This illustrates once more the exceptional role of
the \nnlo{} corrections in this approach.

The leading order terms were evaluated by Eichten {\it et
al.}~\cite{Eichten:1984eu}. The leading $bg$ and $gg$ initiated
processes (\figs{fig::real1}(c) and \ref{fig::gghbb}(a)) were
subsequently added by Dicus and Willenbrock~\cite{Dicus:1989cx}.  Ten
years later, Dicus {\it et al.}~\cite{Dicus:1998hs} (see also
\reference{Balazs:1998sb}) computed the full \nlo{} contribution to
$\bbar\to \higgs$ (and related subprocesses), which leads to the
single logarithmically suppressed term $c_{n1}$ for all $n\geq 0$.
Setting the renormalization and the factorization scale equal to
$\mhiggs$ ($\mu_R=\mu_F=\mhiggs$), they found that the
$\order{\alpha_s}$-corrections to the $\bbar \to \higgs$ subprocess
and the contribution from the tree level $bg\to \higgs b$ subprocess
are quite large, but of opposite sign.  This leads to large
cancellations which are particularly drastic at the \lhc{}. They also
observed that the contribution from $bg\to \higgs b$ becomes
especially large at Higgs boson masses below $\approx 150$\,GeV, meaning
that logarithmically suppressed terms become important in this region.

\FIGURESTAR[ht]{
\begin{tabular}{ccc}
\epsfxsize=4cm
\epsffile[150 500 410 670]{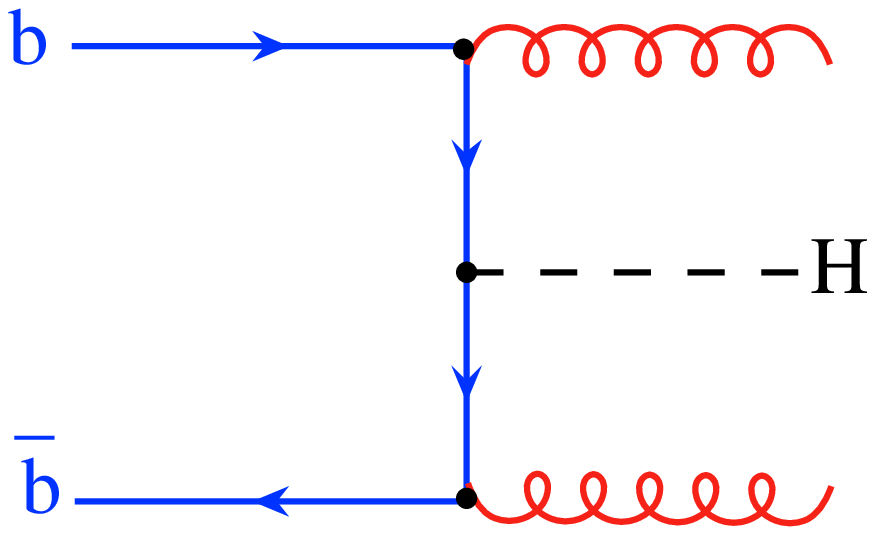} &
\epsfxsize=4cm
\epsffile[150 500 410 670]{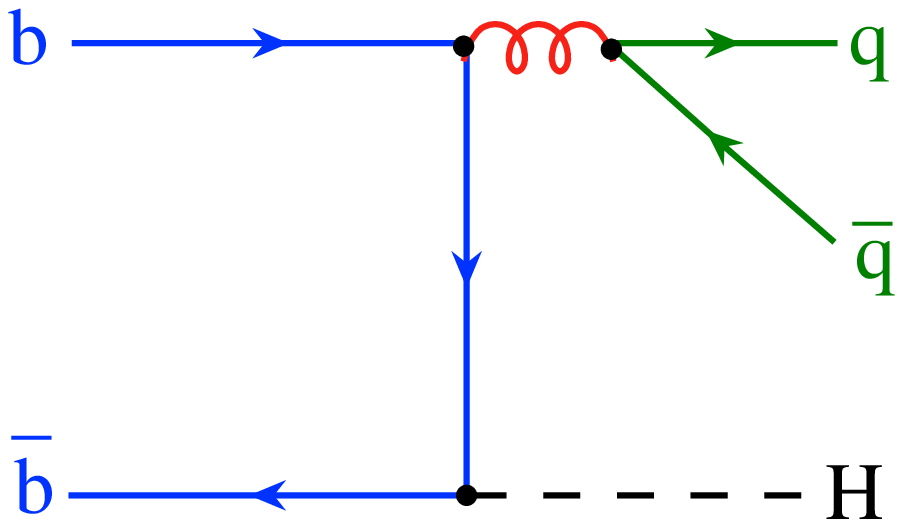} &
\epsfxsize=4.6cm
\epsffile[105 520 380 680]{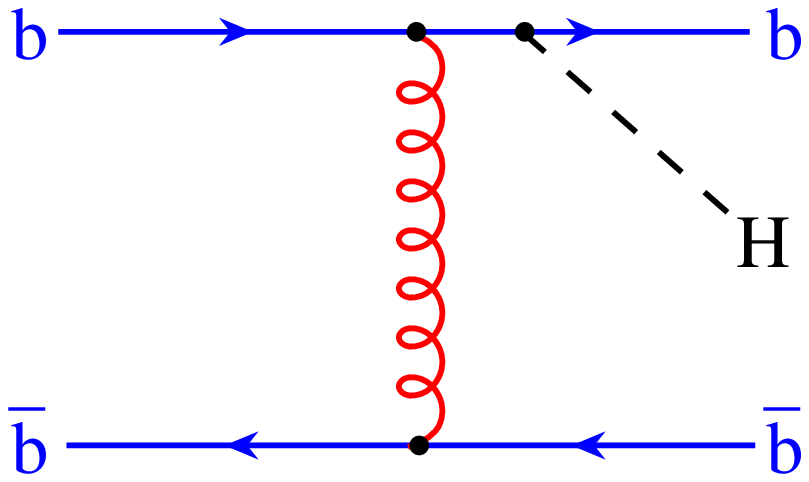} \\
(a) & (b) & (c)\\[2em]
\epsfxsize=4cm
\epsffile[150 500 410 670]{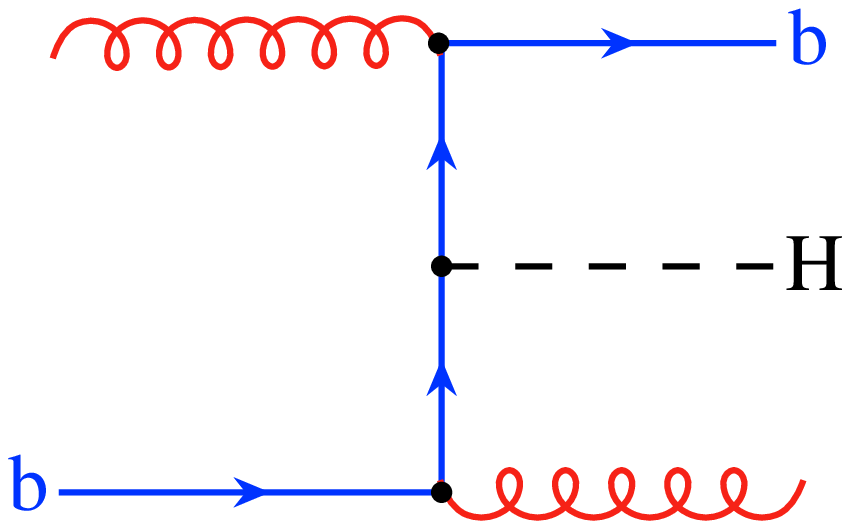} &
\epsfxsize=4.6cm
\epsffile[105 520 380 680]{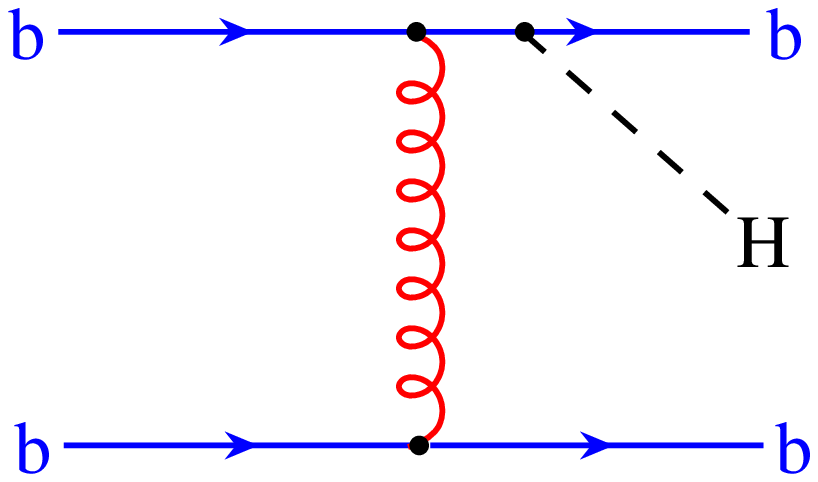} &
\epsfxsize=4.6cm
\epsffile[105 520 380 680]{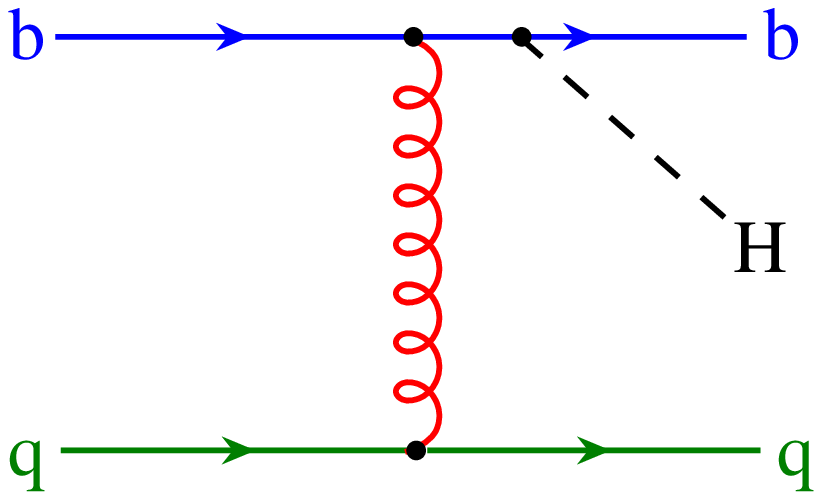} \\
(d) & (e) & (f)
\end{tabular}
\caption[]{\label{fig::real2}Diagrams contributing at \nnlo{}.
  Note that the Higgs boson can couple to the $b$--quarks at any point;
  only representative diagrams are shown.}
}


Recently, Maltoni {\it et al.}~\cite{Maltoni:2003pn} revisited the
\nlo{} calculation in the light of \reference{Plehn:2002vy}, which gives
an argument for the proper choice of the factorization scale when using
the bottom quark density approach. Following that argument,
they determined the factorization scale for the
$(\bbar)\higgs$ process to be $\mu_F\approx\mhiggs/4$. With this choice,
both the \nlo{} corrections from the $gb$ and the $\bbar$ initiated
process turn out to be very well behaved.

As we will show in \sct{sec:results}, the behavior of the \nnlo{}
corrections confirms this choice of scale at \nlo{}, in the sense that
the perturbation theory up to \nnlo{} is very well behaved for this
choice.  This supports the method of
\refs{Plehn:2002vy,Maltoni:2003pn} (see also \reference{Boos:2003yi})
for determining the factorization scale in the \vfs{} approach at
lower orders.  For the process under consideration, it turns out that
the dependence on the unphysical scales of the \nnlo{} result is so
weak that the discussion on the proper scale choice becomes
irrelevant.  The overall conclusion is that the prediction for Higgs
boson production in bottom quark fusion is now under very good
control.


\section{Outline of the calculation}\label{sec:outline}
As discussed before, we will neglect the bottom quark mass everywhere
except in the Yukawa couplings.  The calculation is thus completely
analogous to, say, Drell-Yan production of virtual
photons~\cite{Hamberg:1991np,Harlander:2002wh}: One evaluates virtual
and real corrections to Higgs production in $\bbar$, $gb$, $gg$, $bb$,
$qb$ and $\qqbar$ scattering (and the charge conjugated processes)
and then performs ultraviolet renormalization and mass factorization.

The subprocesses to be evaluated at the partonic level are given as
follows ($q \in\{ u,d,c,s\}$):\\[.5em]
\begin{tabular}{ll}
$\bullet$
 up to two loops:& $\bbar\to \higgs $ \quad
 [\fig{fig::real1}(a)]\\[.2em]
$\bullet$
 up to one loop:& $\bbar\to \higgs g$,\quad $gb\to \higgs b$ \quad
  [\figs{fig::real1}(b), \ref{fig::real1}(c)]\\[.2em]
$\bullet$
 at tree level:&
  \begin{minipage}[t]{35em}
     $\bbar\to \higgs gg$, \quad $\bbar \to \higgs q\bar q$,\\[.2em]
    $\bbar\to \higgs \bbar$,\quad $gb\to \higgs gb$,\\[.2em]
    $bb\to \higgs bb$,\quad $bq\to \higgs bq$\quad
    [\figs{fig::real2}(a)--\ref{fig::real2}(f)]\\[.2em]
     $gg\to \higgs \bbar$\quad
    [\figs{fig::gghbb}(a)--\ref{fig::gghbb}(c)],\\[.2em]
    $q\bar q\to \higgs \bbar$\quad [\fig{fig::gghbb}(d)]\\[.2em]
  \end{minipage}
\end{tabular}
We compute the two-loop virtual terms by employing the method of
\refs{Baikov:2000jg,Harlander:2000mg}, which maps them onto
three-loop two-point functions. In this way, they can be reduced to a
single master integral, using the reduction formulas given in
\refs{Chetyrkin:1981qh,Larin:1991fz}. The master integral has
been computed in \reference{Gonsalves:1983nq}.  The pole parts of this
``$\bbar \higgs$ form factor'' can be compared to the general formula
of \refs{Catani:1998bh,Sterman:2002qn} which provides a welcome
check.  For the generation of the diagrams we use {\abbrev
QGRAF}~\cite{Nogueira:1993ex} as embedded in the automated system {\abbrev
GEFICOM}~\cite{Geficom,Steinhauser:2002rq} (see also
\reference{Harlander:1998dq}).

The one-loop single emission processes are obtained by computing
analytically the full one-loop amplitudes, which are then interfered
with the amplitudes of the tree-level processes.  The two-particle
phase space integrals are also computed analytically.

For the tree-level double emission processes, we express the matrix
elements and phase space measures in terms of the variable
$x=\mhiggs^2/\hat s$, where $\hat s$ is the center of mass energy. Then
we expand the integrands in terms of $(1-x)$~\cite{Harlander:2002wh}.
This leaves us with only one nontrivial phase space integral,
independent of the order of the expansion.  The regular integrands and
the finite integration region ensure the validity of the interchange of
integration and expansion. Keeping of the order of ten terms in the
expansion in $(1-x)$ leads to a hadronic result that is already
phenomenologically equivalent to the analytic result.  By evaluating the
expansion up to sufficiently high orders, however, one can invert the
series~\cite{Kilgore:2002sk} by mapping the expansion onto a set of
basis functions. The latter can be deduced from the known \nnlo{}
Drell-Yan result~\cite{Hamberg:1991np,Harlander:2002wh}.

All algebraic manipulations are performed with the help of the program
{\abbrev FORM}~\cite{Vermaseren:2000nd}.

For a consistent treatment of the \nnlo{} process, it is not
sufficient to evaluate only the partonic cross section at
\nnlo{}. Another ingredient is the proper parton densities, obeying
\nnlo{} Dokshitzer-Gribov-Lipatov-Altarelli-Parisi (\dglap)
evolution. At present, only approximate evolution kernels are known,
derived from moments of the structure
functions~\cite{Retey:2000nq,Larin:1994vu,Larin:1997wd}.  On this
basis, approximate \nnlo{} parton distribution sets have been
evaluated~\cite{Martin:2002dr}. We use this set in all of our
numerical analyses below.  Once parton distributions that use exact
\nnlo{} evolution become available, it is a straightforward task to
update the analysis using the partonic results presented in
Appendix~\ref{sec:partresults}.

Let us now turn to the underlying interaction and the renormalization of the
partonic results.  We ignore the bottom quark mass and the
electroweak interactions, so for our purposes, the Lagrangian is:
\begin{equation}
\begin{split}
  {\cal L}_{b\bar{b}\higgs} &= -\frac{1}{4}F^a_{\mu\nu}F^{a\,\mu\nu}
     + \sum_{q} \bar q\,i\feynsl{D}q
     + \bar{b}\,i\feynsl{D}b -\lambda_b^\bare \bar{b}\higgs{}b\,,
\label{eq::lagbbh}
\end{split}
\end{equation}
where $F^a_{\mu\nu}$ is the gluon field strength tensor, $D_\mu$ is
the \qcd{} covariant derivative, and the sum runs over the quarks
$u,d,s,c$.  $\lambda_b^\bare$ is a bare bottom Yukawa coupling
constant. In the modified minimal subtraction $\Lx\msbar\Rx$ scheme,
the scalar coupling is renormalized such that\footnote{We refrain from
quoting terms proportional to $\gamma_{\rm E}$ and $\ln4\pi$ that drop
out of $\msbar$-renormalized quantities.}
\begin{equation}
\begin{split}
  \lambda_b^\bare &\equiv \lambda_b\,Z_m(\alpha_s)\,,
     \hskip 2cm \LB\phi = h,H\RB\\
  Z_m(\alpha_s) &=
  1 - \api\frac{1}{\ep}\\ + &
   \left(\api\right)^2\,
   \bigg[\frac{1}{\ep^2}\left(\frac{15}{8} - \frac{n_f}{12}\right)
   + \frac{1}{\ep}\left(-\frac{101}{48}  + 
         \frac{5}{72}\,n_f\right) \bigg]\\
   &+ \order{\alpha_s^3}\,,
\end{split}
\end{equation}
where $\ep=(4-D)/2$ and $D$ is the number of space-time dimensions in
which we evaluate the loop (and phase-space) integrals.
$Z_m(\alpha_s)$ is identical to the quark mass renormalization
constant of \qcd~\cite{Chetyrkin:1997dh,Vermaseren:1997fq}.
Here and in the following, we use the short hand notations
$\lambda_b\equiv \lambda_b^{(n_f)}(\muR)$ and $\alpha_s\equiv
\alpha_s^{(n_f)}(\muR)$ for the $\msbar$-renormalized Yukawa and
strong coupling constants, respectively.  $\muR$ is the
renormalization scale, and $n_f$ is the number of ``active'' quark
flavors. We will set $n_f=5$ in our numerical analyses.

There are (at least) two methods of obtaining the result for
pseudoscalar production.  The first is to replace the Yukawa
interaction term in \eqn{eq::lagbbh} with a pseudoscalar interaction,
\begin{equation}
  \lambda_b^\bare \bar{b}\higgs{}b \longrightarrow
    i\lambda_b^\bare \bar{b}\higgs\gamma_5{}b\,,
  \label{eq::pseudobbh}
\end{equation}
and proceed by direct calculation.

The second method is to exploit the chiral symmetry of the bottom
quarks in \eqn{eq::lagbbh}, which implies that we are free to perform
independent left-handed and right-handed phase rotations of the bottom
quarks.  If we perform the rotation
\begin{equation}
  b_R \to i\,b^\prime_R \hskip 3cm b_L \to b^\prime_L\,,
\end{equation}
the Lagrangian becomes
\begin{equation}
\begin{split}
  {\cal L}_{b\bar{b}\higgs}
     \to& -\frac{1}{4}F^a_{\mu\nu}F^{a\,\mu\nu}
     + \sum_{q} \bar q\,i\feynsl{D}q
     + \bar{b}^\prime\,i\feynsl{D}b^\prime
     -i\lambda_b^\bare \bar{b}^\prime\higgs\gamma_5{}b^\prime\,,
\label{eq::lagbbh2}
\end{split}
\end{equation}
and we find the same interaction Lagrangian as in \eqn{eq::pseudobbh}.
This implies that the cross section for pseudoscalar Higgs boson
production, written in terms of the Yukawa coupling $\lambda_b$,
is identical to the cross section for scalar Higgs boson
production to all orders in $\alpha_s$.

Following the prescription of Larin~\cite{Larin:1993tq}\footnote{Note
that only the e-print of \reference{Larin:1993tq} discusses the
renormalization of the pseudoscalar current.} for the treatment of
$\gamma_5$ in dimensional regularization, we have performed the direct
calculation through \nnlo{} and find that this is indeed the case.

Even in the direct calculation, one can see that this identity will
hold to all orders in $\alpha_s$ with the following argument.  If we
square the amplitude before computing loop integrals, all fermion lines
are closed loops.  The fact that we set the bottom quark mass to zero
means that both Higgs boson vertices (in both the scalar and
pseudoscalar cases) must appear on the same fermion line.  If only
one Higgs vertex were to appear on a fermion line, there would be an
odd number of $\gamma$ matrices in the fermion trace which would
therefore vanish.  In the pseudoscalar case, this means that
nonvanishing fermion traces must contain either zero or two
$\gamma_5$ matrices.  The prescription of Larin~\cite{Larin:1993tq}
allows one to assume anticommutativity of $\gamma_5$ and identify
$\gamma_5^2=1$ when two $\gamma_5$-matrices are on the same fermion
line.  Thus, the $\gamma_5$-matrices can be eliminated and we see that
the calculation for pseudoscalar Higgs boson production is identical,
diagram by diagram of the squared amplitude, to that for scalar Higgs
boson production, apart from the different Yukawa couplings.

For the sake of completeness, let us remark that the Standard Model
value for the coupling constant is given by $\lambda_{b} =
\sqrt{2}m_b/v$, where $v\approx 246$\,GeV is the vacuum expectation
value for the Higgs boson field, and $m_b$ is the running $\msbar$ mass of
the bottom quark, $m_b(\mu_R)$, evaluated at the renormalization scale
$\muR$.  In the \mssm{} we have
\begin{equation}
\begin{split}
\lambda_b &= \left\{
  \begin{array}{ll}
    \displaystyle
-\sqrt{2}\frac{m_b}{v}\frac{\sin\alpha}{\cos\beta}\,,
   &\hskip 2cm \phi = h\,,\\[15pt]
    \displaystyle
 \phantom{-}\sqrt{2}\frac{m_b}{v}\frac{\cos\alpha}{\cos\beta}\,,
   &\hskip 2cm \phi = H\,,\\[15pt]
    \displaystyle
 \phantom{-}\sqrt{2}\frac{m_b}{v}\tan\beta\,,
   &\hskip 2cm \phi = A\,.
 \end{array}
\right.
   \label{eq::mssmyukawa}
\end{split}
\end{equation}

The renormalized partonic results have a dependence on the unphysical
scales $\mu_F$ and $\mu_R$, both explicitly in terms of logarithms,
and implicitly through the parameters $\alpha_s(\mu_R)$ and
$\lambda_b(\mu_R)$.  The variation of $\alpha_s$ and $\lambda_b$ with
$\mu_R$ is governed by the renormalization group equations
(\rge{}s)
\begin{equation}
\mu_R^2\frac{\dd}{\dd\mu_R^2} a_s =
\beta(a_s)\,a_s\,,\quad
\mu_R^2\frac{\dd}{\dd\mu_R^2}\lambda_b =
\gamma^m(a_s)\,\lambda_b\,,
\quad a_s\equiv \frac{\alpha_s}{\pi}\,,
\label{eq::rgalpha}
\end{equation}
where
\begin{equation}
\begin{split}
\beta(a_s) &= -a_s\beta_0 - a_s^2\beta_1 - a_s^3\beta_2
   + \order{a_s^4}\,,\\[.75em]
\beta_0 &= \frac{11}{4} - \frac{1}{6}\,n_f\,,\\[.75em]
\beta_1 &= \frac{51}{8} - \frac{19}{24}\,n_f\,,\\[.75em]
\beta_2 &= \frac{2857}{128} - \frac{5033}{1152}\,n_f
  + \frac{325}{3456}\,n_f^2\,,\\[15pt]
\gamma^m(a_s) &= -a_s\gamma^m_0 - a_s^2\gamma^m_1 - a_s^3\gamma^m_2
   + \order{a_s^4}\,,\\[.75em]
\gamma^m_0 &= 1\,,\\[.75em]
\gamma^m_1 &= \frac{101}{24} - \frac{5}{36}\,n_f\,,\\[.75em]
\gamma^m_2 &= 
  \frac{1249}{64} - \left(\frac{277}{216} + \frac{5}{6}\,\zeta_3\right)\,n_f
  - \frac{35}{1296}\,n_f^2\,.
\label{eq::betgam}
\end{split}
\end{equation}
Here, $\zeta_n\equiv \zeta(n)$ is Riemann's $\zeta$-function ($\zeta_3
\approx 1.20206$).  In order to evaluate $\alpha_s(\mu_R)$ from the
initial value\footnote{The numerical value of $\alpha_s(M_Z)$ has to
be set in accordance with the parton sets that are used, see below.}
$\alpha_s(M_Z)$, $\beta(a_s)$ is expanded up to $\alpha_s^\ell$, with
$\ell=1$ at \lo{}, $\ell=2$ at \nlo{}, and $\ell=3$ at \nnlo{}.  The
resulting differential equation of \eqn{eq::rgalpha} is solved
numerically.

In order to evaluate $\lambda_b(\mu)$ from its initial value
$\lambda(\mu_0)$, we combine the two \rge{}s of \eqn{eq::rgalpha}
to obtain
\begin{equation}
\begin{split}
\lambda_b(\mu) = \lambda_b(\mu_0)\,\frac{c(a_s(\mu))}{c(a_s(\mu_0))}\,,
\end{split}
\end{equation}
with
\begin{equation}
\begin{split}
c(a) = a^{c_0}\bigg\{
&1 + (c_1 - b_1c_0)\, a\\&
+ \frac{1}{2}\left[ (c_1 - b_1c_0)^2 + c_2 - b_1c_1 + b_1^2c_0 -
  b_2c_0\right]a^2\\
 &+ \order{a^3}\bigg\}\,,\qquad\qquad
c_i \equiv \frac{\gamma_{i}^m}{\beta_0}\,,
\qquad
b_i \equiv \frac{\beta_i}{\beta_0}\,.
\label{eq::cx}
\end{split}
\end{equation}

\FIGURE[ht]{
  \begin{tabular}{c}
    \epsfxsize=8.cm
    \epsffile{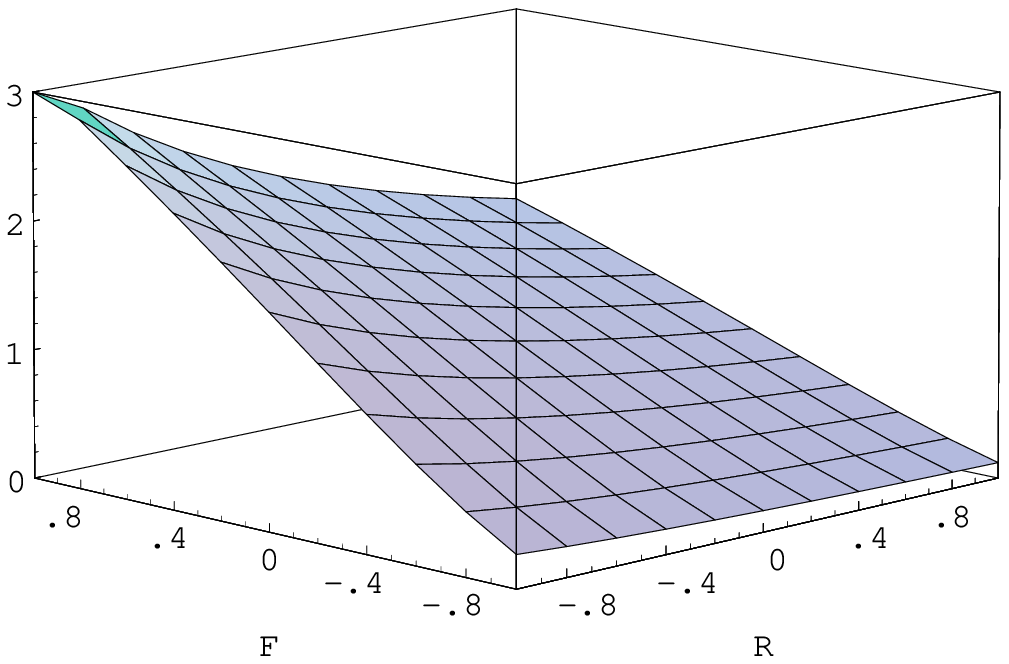} \\
    (a) \\[1em]
    \epsfxsize=8.cm
    \epsffile{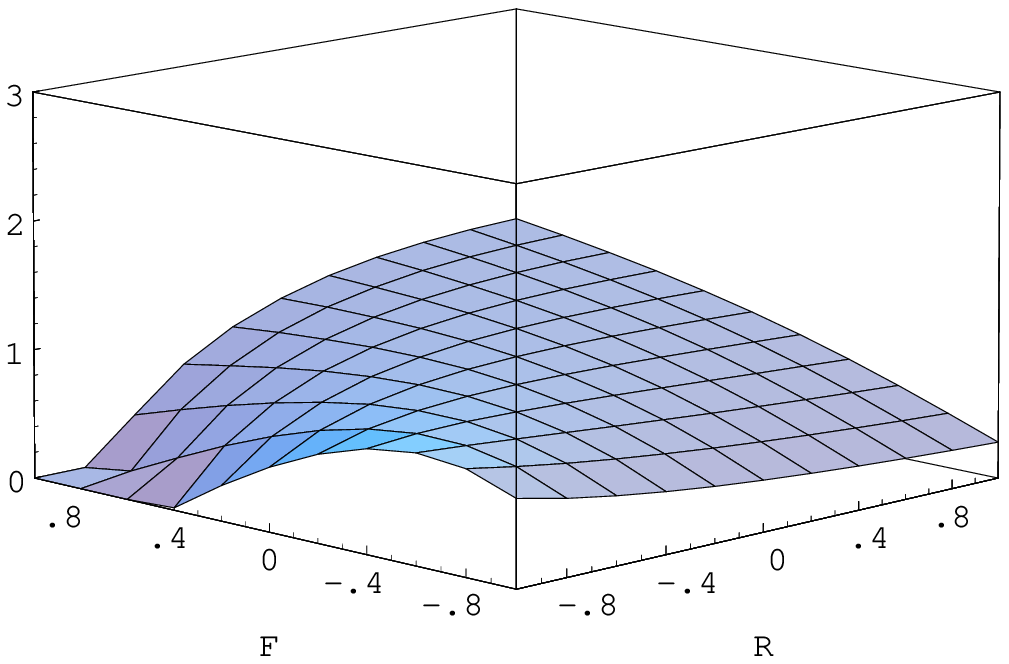} \\
    (b) \\[1em]
    \epsfxsize=8.cm
    \epsffile{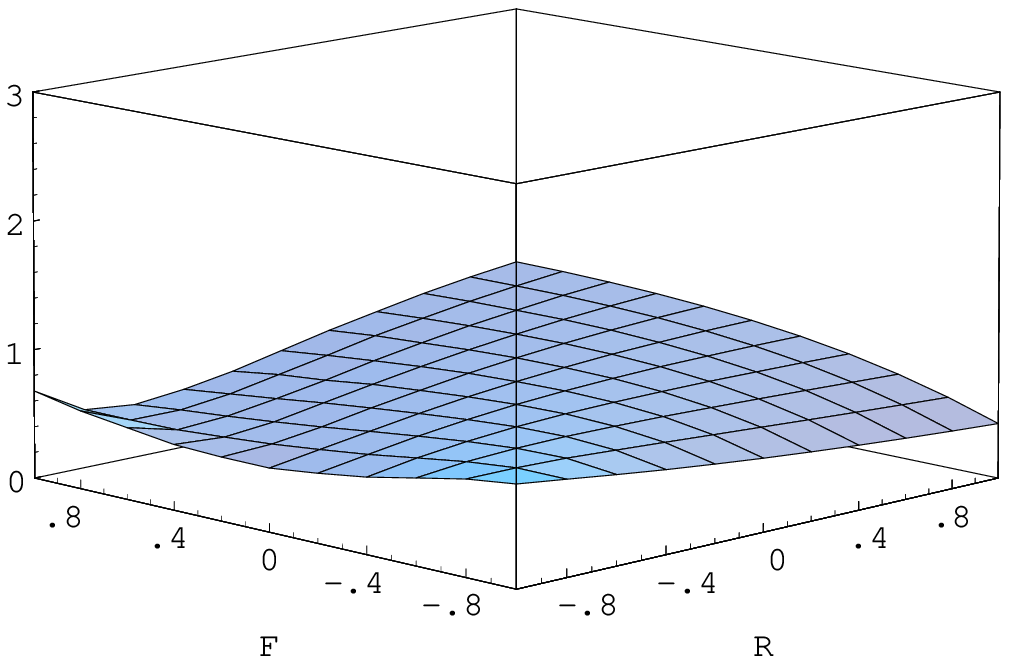} \\
    (c)
  \end{tabular}
    \caption[]{\label{fig::plo}\sloppy
      The cross section $\sigma(pp\to (\bbar)H+X)$ (in picobarns) at
      (a) \lo{}, (b) \nlo{}, (c) \nnlo{} for the \lhc{}.  The axes
      labels are $F=\log_{10}(\muF/M_H)$ and $R=\log_{10}(\muR/M_H)$.
      Thus, the point $\muR=M_H$, $\muF=0.25\,M_H$ corresponds to $R=0$,
      $F=-0.6$.  The Higgs boson mass is set to $M_H=120$\,GeV.
      }}


\FIGURE[ht]{
  \begin{tabular}{c}
    \epsfxsize=8.cm
    \epsffile{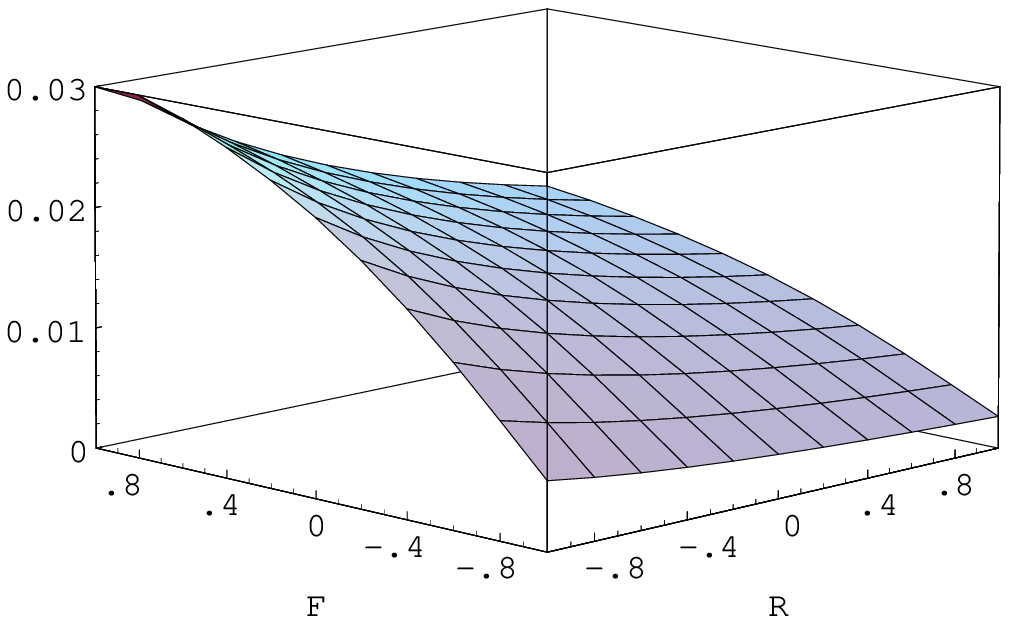} \\
    (a) \\[1em]
    \epsfxsize=8.cm
    \epsffile{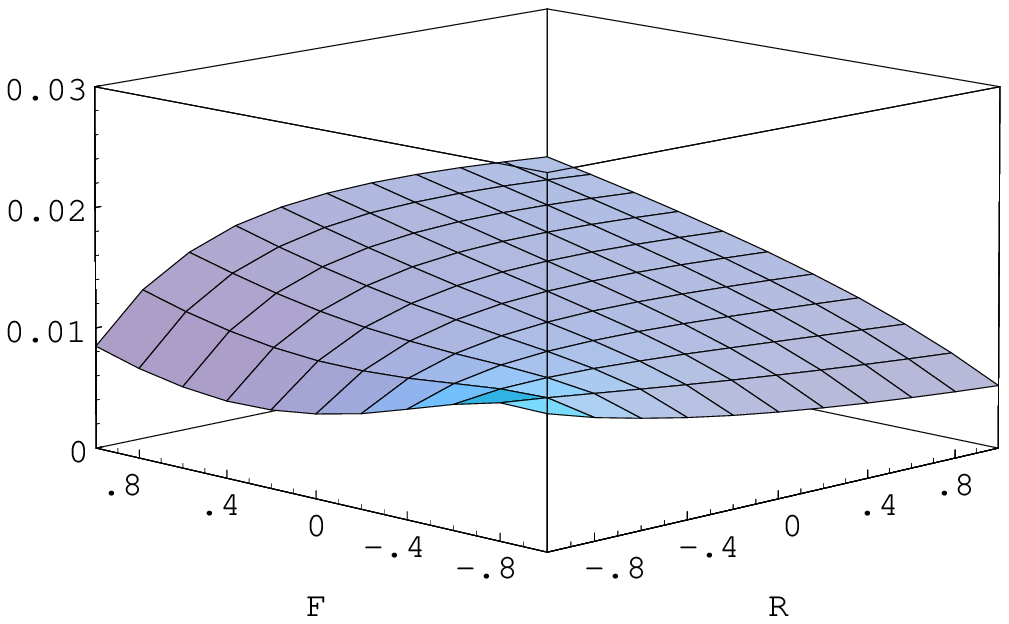} \\
    (b) \\[1em]
    \epsfxsize=8.cm
    \epsffile{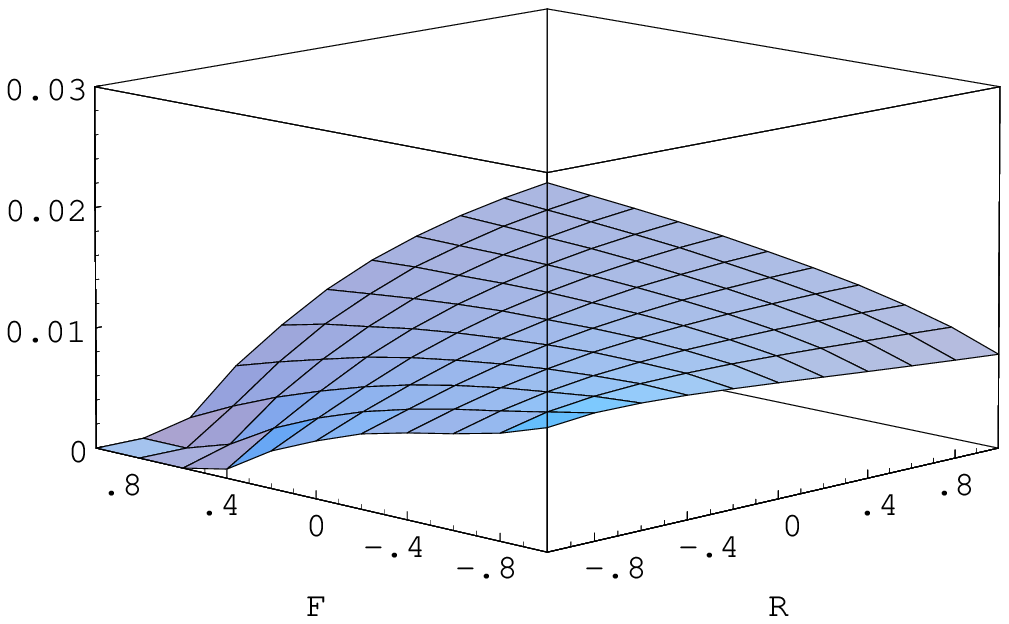} \\
    (c)
  \end{tabular}
    \caption[]{\label{fig::plo_tev}\sloppy
      The cross section $\sigma(p\bar p\to (\bbar)H+X)$ (in picobarns)
      at (a) \lo{}, (b) \nlo{}, (c) \nnlo{} for
      $\sqrt{s}=1.96$\,GeV.  The notation is the same as in \fig{fig::plo}.
      The Higgs boson mass is set to $M_H=120$\,GeV.  }}


$\beta_i^m$ and $\gamma_i$ have been defined in \eqn{eq::betgam}.  Both
$a_s(\mu)$ and $a_s(\mu_0)$ are calculated from $\alpha_s(M_Z)$ using
the procedure described above.  Working at \lo{} (\nlo{}, \nnlo{}), we
truncate the term in braces at order $a^0$ ($a^1$, $a^2$).

Convolution of the partonic cross section with the parton densities
cancels the $\mu_F$ dependence up to higher orders and results in the
physical hadronic cross section. The variation of the hadronic cross
section with $\mu_F$ and $\mu_R$ is thus an indication of the size of
higher order effects.


\section{Results}\label{sec:results}
%

\FIGURE[ht]{
  \begin{tabular}{c}
    \epsfxsize=8.cm
    \epsffile[110 265 465 560]{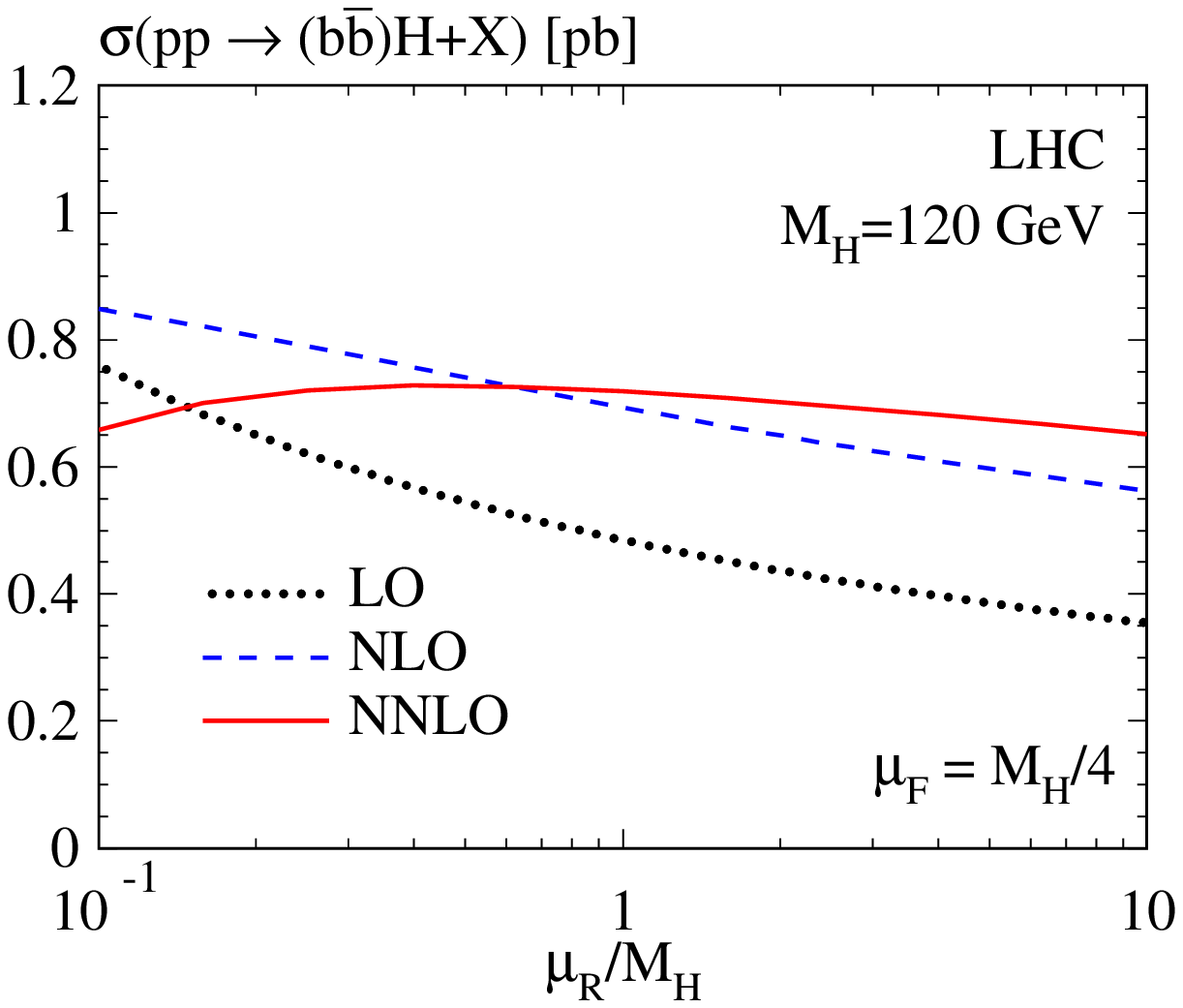} \\
    (a) \\
    \epsfxsize=8.cm
    \epsffile[110 265 465 560]{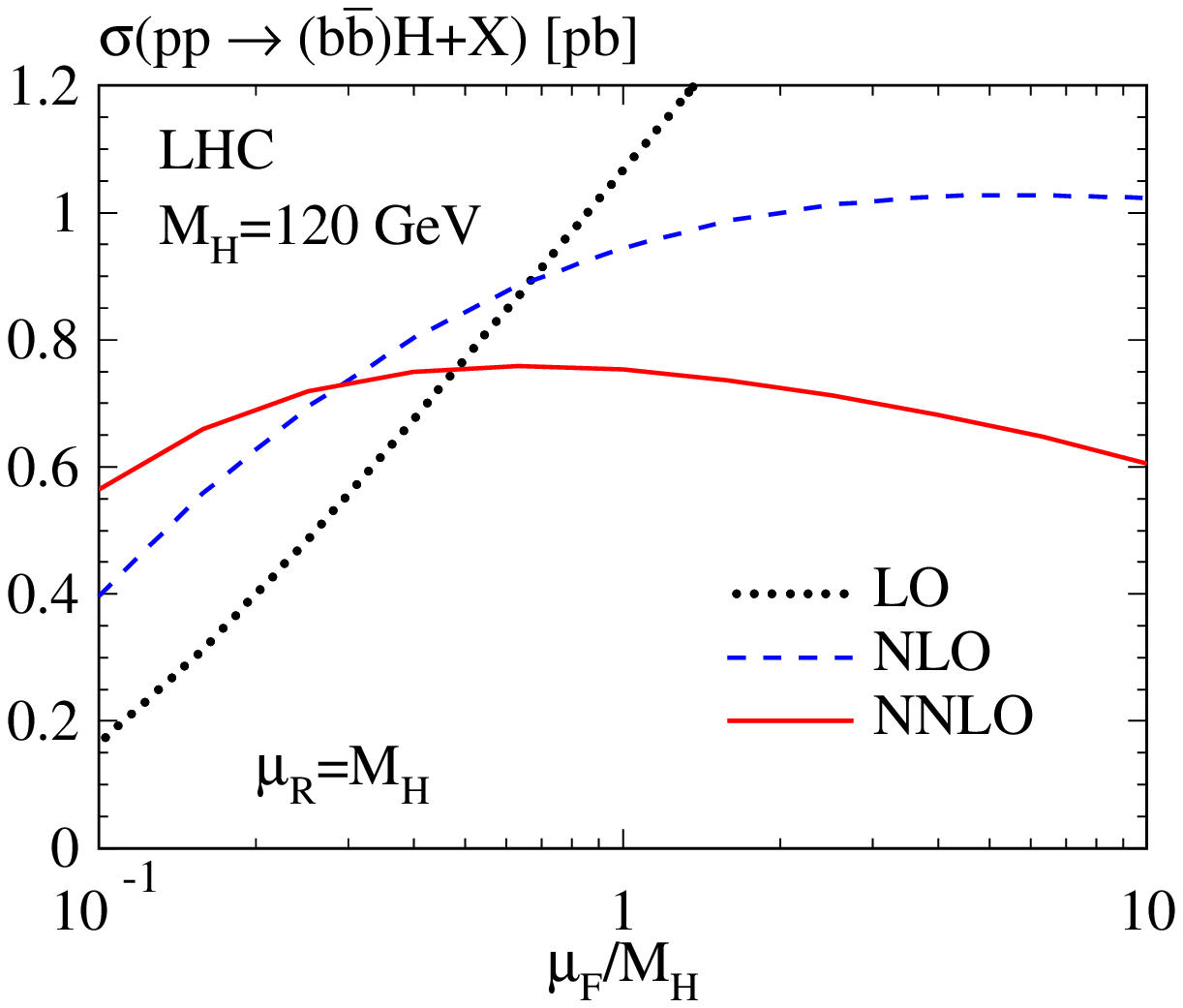} \\
    (b)
  \end{tabular}
    \caption[]{\label{fig::murf14a}\sloppy
      Cross section for $pp\to (\bbar)H+X$ at $\sqrt{s}=14$\,TeV,
      $M_H=120$\,GeV.  (a) $\muR$ dependence for $\muF=0.25M_H$; (b)
      $\muF$ dependence for $\muR=M_H$.  }}


\FIGURE[ht]{
  \begin{tabular}{c}
    \epsfxsize=8.cm
    \epsffile[110 265 465 560]{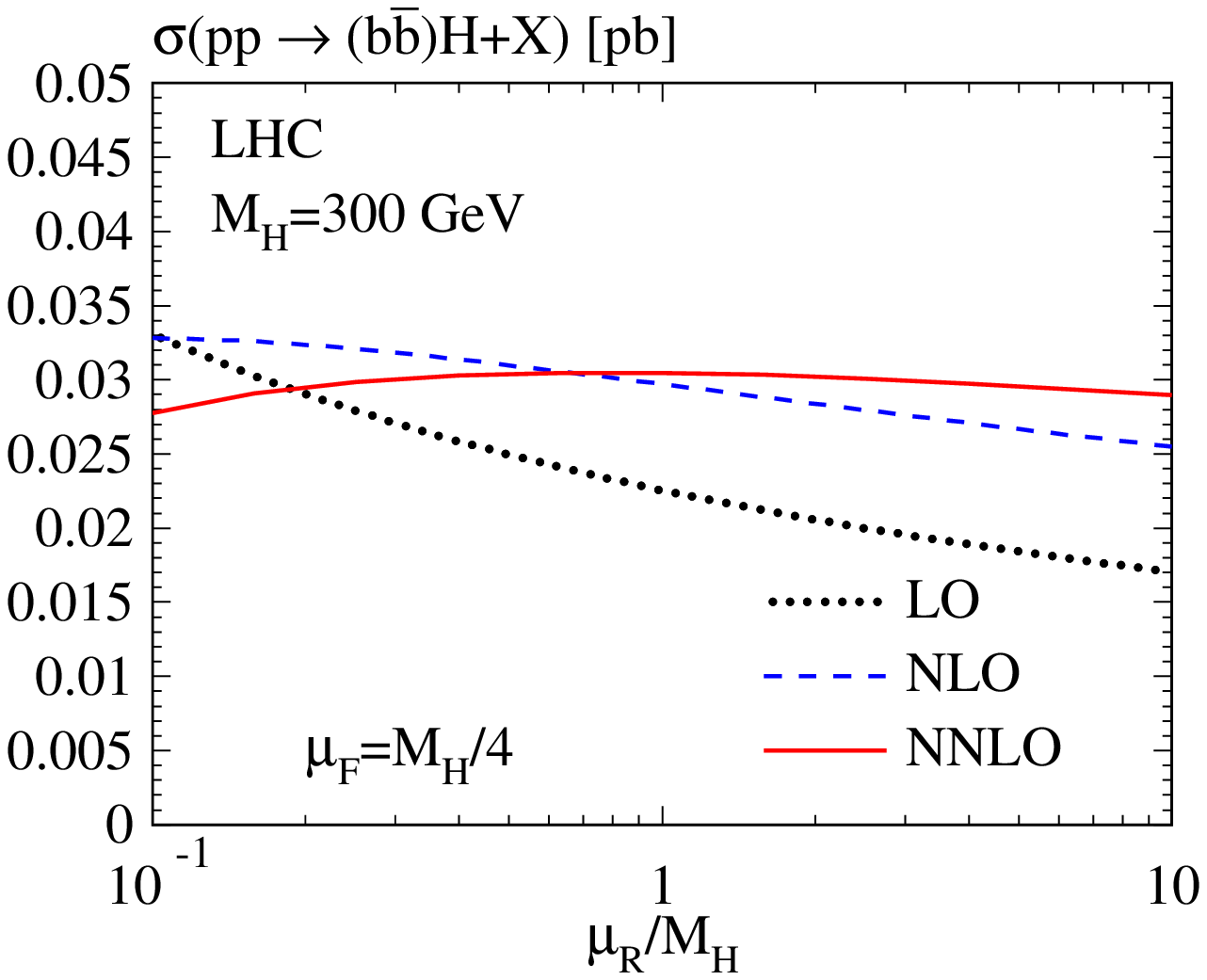} \\
    (a) \\
    \epsfxsize=8.cm
    \epsffile[110 265 465 560]{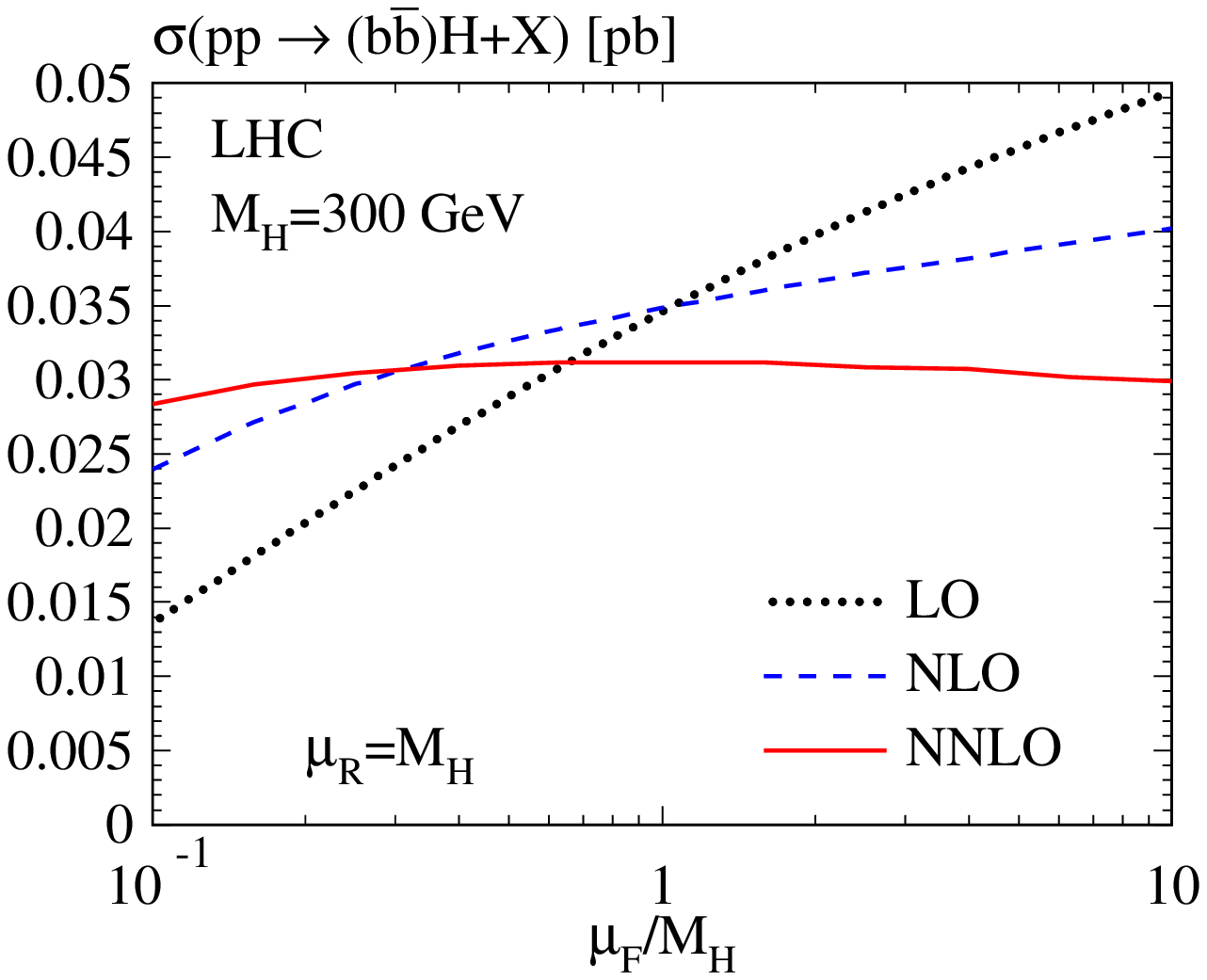} \\
    (b)
  \end{tabular}
    \caption[]{\label{fig::murf14b}\sloppy
      Cross section for $pp\to (\bbar)H+X$ at $\sqrt{s}=14$\,TeV,
      $M_H=300$\,GeV.  (a) $\muR$ dependence for $\muF=0.25M_H$; (b)
      $\muF$ dependence for $\muR=M_H$.  }}


The analytic expressions for the partonic cross section are quite
voluminous and will be deferred to the appendix.  In this section, we
study the behavior of the \nnlo{} result with respect to variations of
the input parameters, in particular the Higgs boson mass and the
collider type (\lhc{} and Tevatron). Special emphasis is placed on the
variation of the results with the renormalization and factorization
scale, from which we estimate the theoretical uncertainty of the
prediction for Higgs boson production in $\bbar$ annihilation.

Because the cross sections for the neutral Higgs bosons in $\bbar$
annihilation differ only in the magnitudes of the Yukawa couplings
(within our approximations), we will restrict our discussion to the
production of a Standard Model Higgs boson.  In the limit that
supersymmetric partners are heavy, their virtual contributions are
insignificant and the predictions for supersymmetric Higgs bosons can
be obtained from the Standard Model values by rescaling them with the
proper coupling constants (cf.\ \eqn{eq::mssmyukawa}).

All the numerical results have been obtained using
Martin-Roberts-Stirling-Thorne ({\abbrev MRST}) parton distributions.
In particular, we use the {\abbrev MRST2001} sets~\cite{Martin:2001es}
at \lo{} $\LB\alpha_s(M_Z)=0.130\RB$ and \nlo{} $\LB\alpha_s(M_Z)=0.119\RB$,
and {\abbrev MRSTNNLO}~\cite{Martin:2002dr} at \nnlo{}
$\LB\alpha_s(M_Z)=0.1155\RB$.


\FIGURE[ht]{
  \begin{tabular}{c}
    \epsfxsize=8.cm
    \epsffile[110 265 465 560]{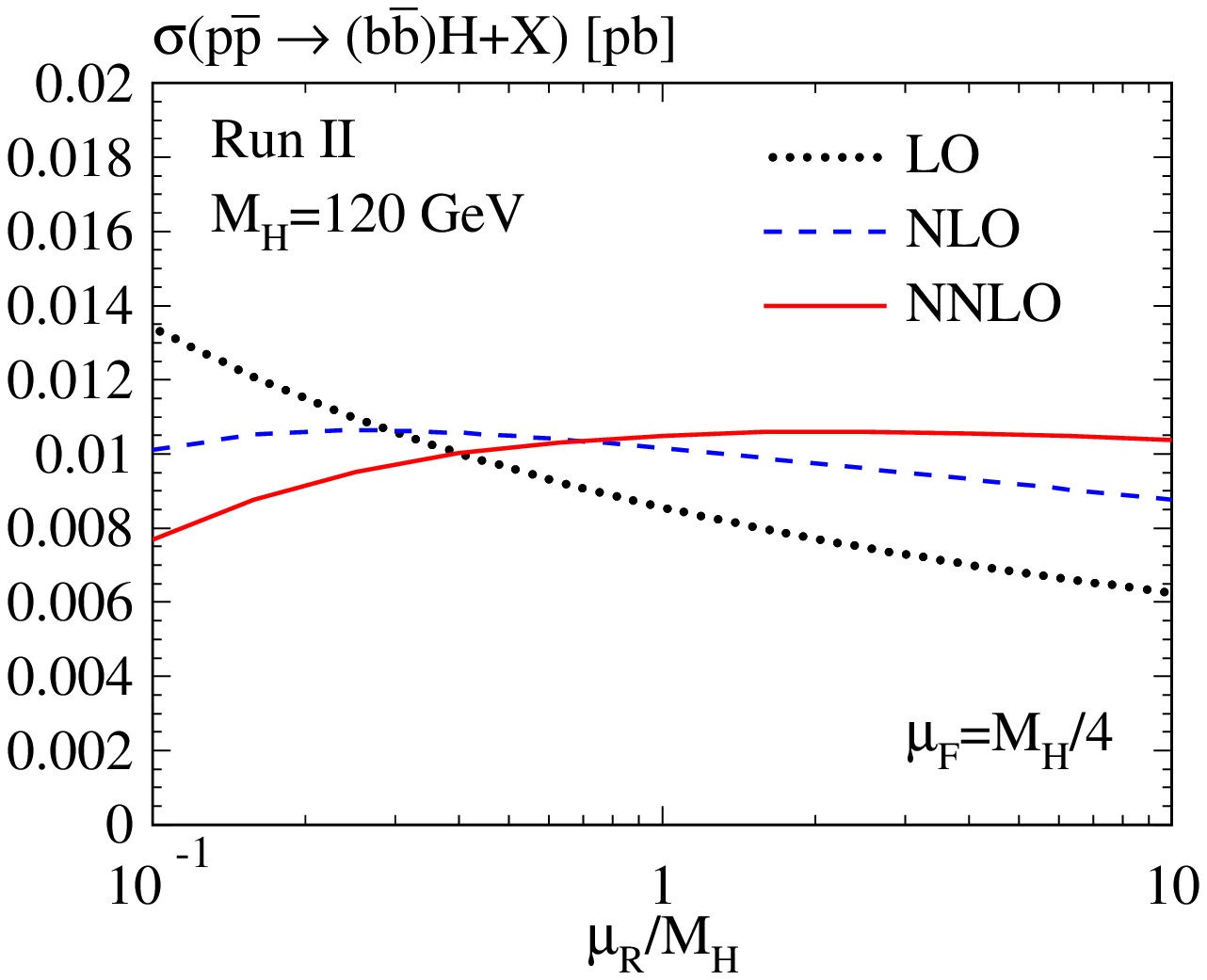} \\
    (a) \\
    \epsfxsize=8.cm
    \epsffile[110 265 465 560]{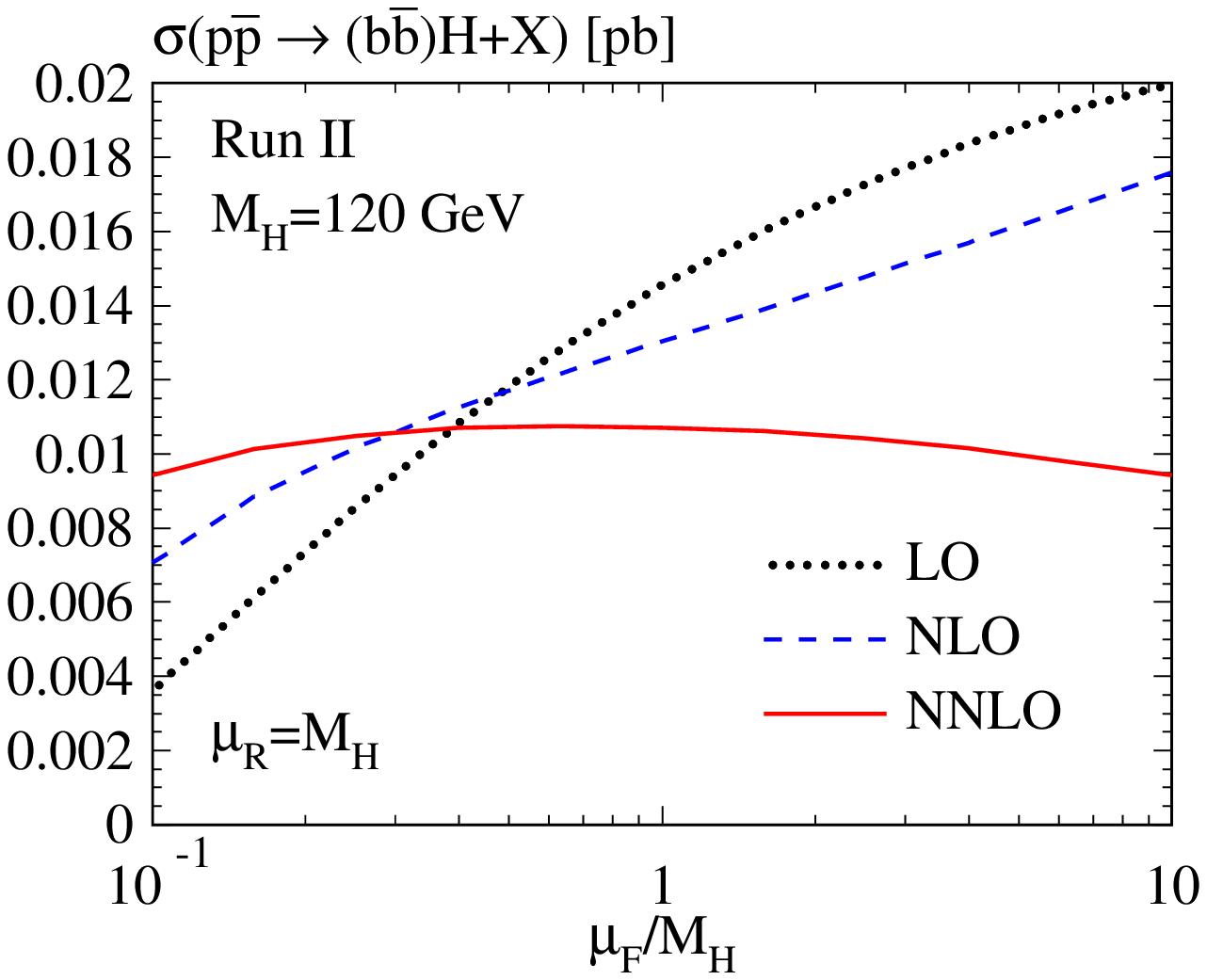} \\
    (b) 
  \end{tabular}
    \caption[]{\label{fig::murf2a}\sloppy
      Cross section for $p\bar p\to (\bbar)H+X$ at $\sqrt{s}=1.96$\,TeV,
      $M_H=120$\,GeV.  (a) $\muR$ dependence for $\muF=0.25M_H$; (b)
      $\muF$ dependence for $\muR=M_H$.  }}



\FIGURE[ht]{
  \begin{tabular}{c}
    \epsfxsize=8.cm
    \epsffile[110 265 465 560]{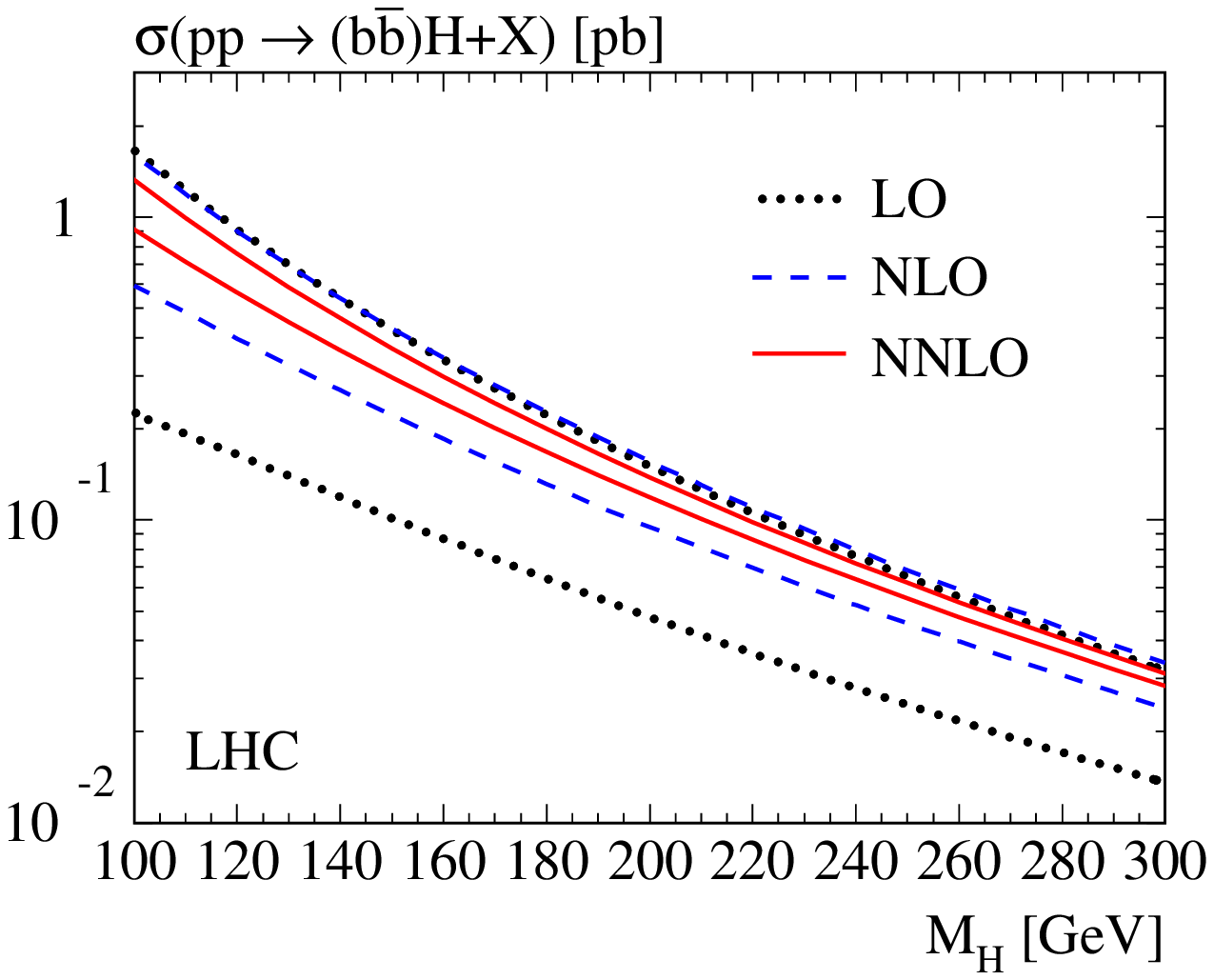}\\
    (a) \\
    \epsfxsize=8.cm
    \epsffile[110 265 465 560]{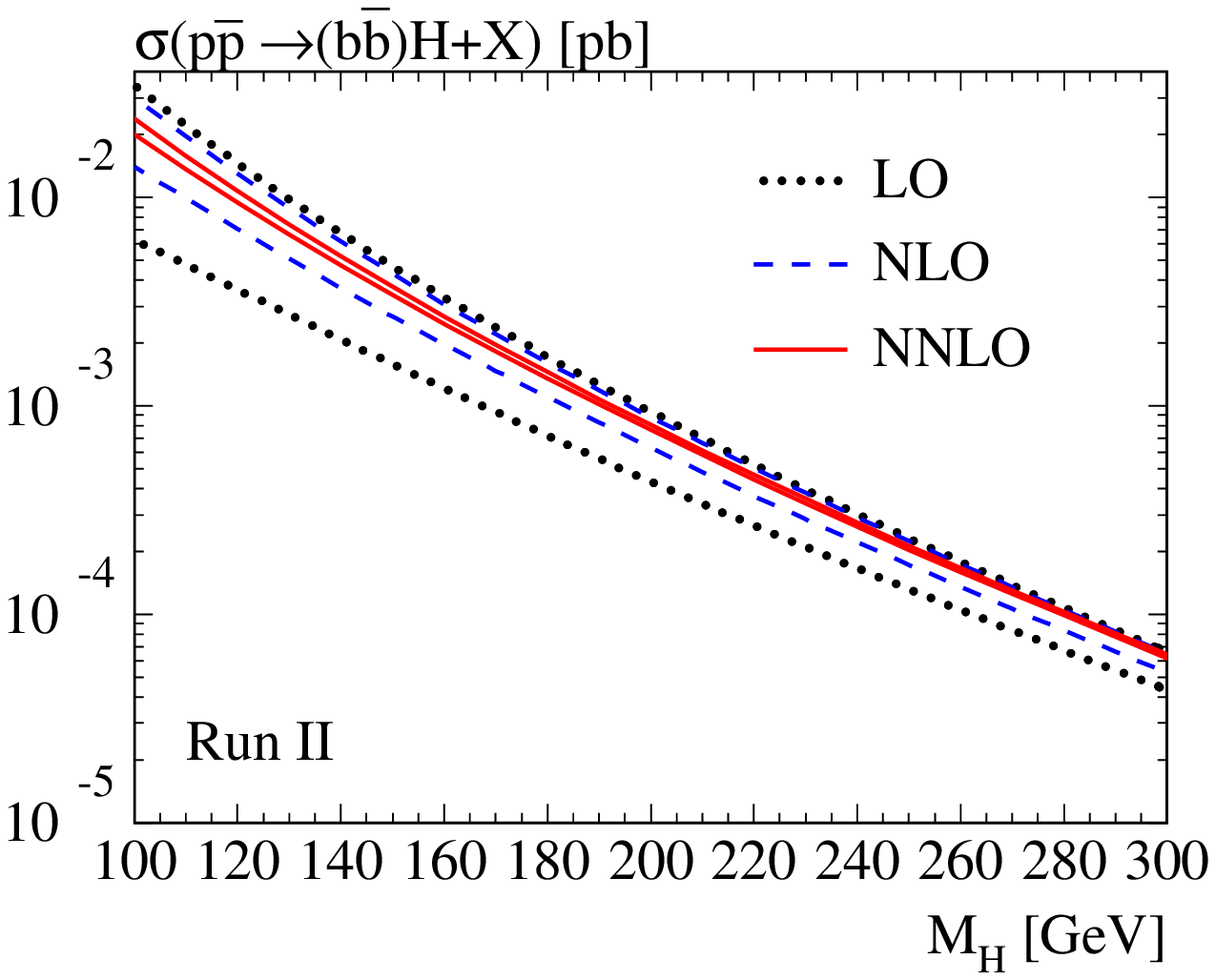}\\
    (b)
  \end{tabular}
    \caption[]{\label{fig::bbhnnlo}\sloppy
      Cross section for Higgs boson production in bottom quark annihilation
      at (a) the \lhc{} and (b) the Tevatron (Run~II) at \lo{}
      (dotted), \nlo{} (dashed) and \nnlo{} (solid).  The upper
      (lower) line corresponds to a choice of the factorization scale
      of $\muF=0.7\,M_H$ ($\muF=0.1\,M_H$). The renormalization scale
      is set to $\muR=M_H$.   }}


In order to obtain an overall picture of the renormalization and
factorization scale dependence of the cross section, we plot
$\sigma(pp\to (\bbar)H+X)$ for $\sqrt{s}=14$\,TeV as a function of the
two parameters $\muF$ and $\muR$ in \fig{fig::plo}. The corresponding
plot for the Tevatron, i.e., $\sigma(p\bar p\to (\bbar)H+X)$ for
$\sqrt{s}=1.96$\,TeV, is shown in \fig{fig::plo_tev}.  The Higgs boson
mass is fixed to $M_H = 120$\,GeV.  Subpanels (a), (b), and (c) show
the \lo{}, \nlo{}, and \nnlo{} prediction, respectively. Note the
extremely large variation of the scales, by a factor of 10 above and
below $M_H$.  Apart from the region of large $\muF$ and small $\muR$,
one observes a clear reduction of the scale dependence with increasing
order of perturbation theory, both for $\muF$ and $\muR$.  Notably, we
find minimal radiative corrections and a particularly weak dependence
on the renormalization and factorization scales in the vicinity of
$(\muR,\muF)=(\mhiggs,0.25\,\mhiggs) \equiv (\bar \mu_R,\bar \mu_F)$.
This agrees with the observation of \reference{Maltoni:2003pn} that
the proper factorization scale for this process should be around
$\muF=\mhiggs/4$.

To illustrate this observation, we display separately the $\muR$- and
$\muF$-variation of the cross section at the \lhc{} in
\fig{fig::murf14a} for $M_H=120$\,GeV, and in \fig{fig::murf14b} for
$M_H=300$\,GeV. In subpanels (a), the factorization scale is fixed
to $\muF=\bar\mu_F=0.25\,M_H$, and the renormalization scale is varied
within $0.1\leq \muR/M_H\leq 10$.  In subpanels (b), the
renormalization scale is fixed to $\muR=\bar\mu_R=M_H$, and the
factorization scale is varied within $0.1\leq \muF/M_H\leq 10$.  The
reduction in the scale dependence with increasing order of
perturbation theory is clearly visible.  As opposed to the lower order
curves which have a monotonic dependence on $\mu_{R/F}$ within the
displayed range, the \nnlo{} curves develop a maximum, so that it is
possible to define a ``point of least sensitivity'' for them. In all
cases, this falls nicely into a region where the radiative corrections
are small. Note also that the central values for the \nnlo{} curves
are perfectly consistent between panels (a) and (b).  These
observations confirm that $\bar\mu_F=0.25\,M_H$ and $\bar\mu_R=M_H$
are indeed the appropriate scale choices for this process.

The corresponding curves for Run~II at the Tevatron are shown in
Fig.\,\ref{fig::murf2a} (we only show results at the Tevatron for
$M_H=120$\,GeV).  As opposed to the \lhc{}, the reduction of the
renormalization scale dependence with increasing order of perturbation
theory is less drastic. One even observes a slight increase in the
$\muR$ dependence between \nlo{} and \nnlo{}. However, the absolute
variation is very small. The factorization scale dependence is quite
similar to that observed for the \lhc{}. Again, the central values for
the \nnlo{} curves of subpanels (a) and (b) coincide nicely.
Note that the cross section at the Tevatron is typically about two
orders of magnitude smaller than at the \lhc{}.

\fig{fig::bbhnnlo}(a) shows the \lo{}, \nlo{}, and \nnlo{}
predictions for the cross section $\sigma(pp\to (\bbar)H+X)$ at the
\lhc{} as a function of the Higgs boson mass. The two curves at each
order correspond to two different choices of the factorization scale,
$\muF=0.1\,M_H$ and $\muF=0.7\,M_H$. From subpanels (b) of
\figs{fig::murf14a} and \ref{fig::murf14b} one can see that this
roughly defines the maximal $\muF$-variation at \nnlo{} between
$0.1\,M_H$ and $10\,M_H$.  Since the renormalization scale dependence
is very weak [cf.\ subpanels (a) of \figs{fig::murf14a} and
\ref{fig::murf14b}], we fix $\muR=\bar\mu_R=M_H$.  Taking the width of
these bands as an indication of the theoretical uncertainty, we
observe an improvement of the accuracy of the prediction for
$M_H=120$\,GeV from $70\%$ at \lo{} to $40\%$ at \nlo{} to $15\%$ at
\nnlo{}. At larger Higgs boson masses, the scale uncertainty is smaller,
amounting to $40\%$ at \lo{}, $17\%$ at \nlo{}, and $5\%$ at \nnlo{}
for $M_H=300$\,GeV.

The cross section for the Tevatron at $\sqrt{s}=1.96$\,TeV
center-of-mass energy is shown in \fig{fig::bbhnnlo}(b).  Here the
renormalization scale dependence within the range $0.1\leq
\muR/M_H\leq 10$ at \nnlo{} is larger than the factorization scale
dependence (cf.\,Fig.\,\ref{fig::murf2a}). Nevertheless, we apply the
same prescription as for the \lhc{} and plot the \lo{}, \nlo{}, and
\nnlo{} cross section at $(\muR,\muF)=(1,0.1)\,M_H$ and
$(\muR,\muF)=(1,0.7)\,M_H$. This is justified since $\muR$-variation
on absolute scales is still very small, in particular if it is
restricted to a more reasonable range of about a factor of five above
and below $M_H$.  We obtain a reduction of the scale uncertainty at
the Tevatron for $M_H=120$\,GeV from around $60\%$ at \lo{}, to $30\%$
at \nlo{}, to $10\%$ at \nnlo{}.


\section{Conclusions}

We have computed the total cross section for Higgs boson production in
$\bbar$ fusion at \nnlo{} in \qcd{}. We have argued that the \nnlo{}
plays an exceptional role in this process, as it incorporates all
subleading logarithms at order $\alpha_s^2$.  The results are very
stable with respect to changes of the renormalization and
factorization scales. We find that the radiative corrections are
particularly small at factorization scales of around $\muF=\mhiggs/4$,
in agreement with the arguments of \refs{Plehn:2002vy,Maltoni:2003pn}.

We conclude that the inclusive cross section for Higgs boson
production in bottom quark annihilation at hadron colliders is under
good theoretical control.

\paragraph*{Acknowledgments.}
We would like to thank Scott Willenbrock, Fabio Maltoni, and Zack
Sullivan for their encouragement and valuable comments.  We are
grateful to Tilman Plehn and Werner Vogelsang for enlightening
discussions on bottom quark parton densities.  R.V.H. would like to
thank Andr\'e Turcot for emphasizing the importance of this process,
and Kostia Chetyrkin for useful comments.  The work of W.B.K. was
supported by the U.~S.~Department of Energy under Contract
No.~DE-AC02-98CH10886.


\begin{appendix}
\section{Partonic results}\label{sec:partresults}

It is convenient to write the partonic cross section in the following
way:
\begin{equation}
\begin{split}
  \hat\sigma_{ij}(x) &= \sigma^0\,\Delta_{ij}(x)\,,
     \qquad i,j \in \{b,\bar{b},g,q,\bar{q}\}\,,
\end{split}
\end{equation}
where $\hat\sigma_{ij}$ is the cross section for the process $ij \to
\phi+X$. $i$ and $j$ label the partons in the initial state, $\phi$
means either a scalar or pseudoscalar Higgs boson, and $X$ denotes
any number of quarks or gluons in the final state.  Here and in what
follows, $q$ denotes any of the light quarks $u,d,s,c$.  The
normalization factor, $\sigma_0$, is
\begin{equation}
\sigma_0 = \frac{\pi}{12}\frac{\lambda_b^2}{M_\phi^2}\,.
\end{equation}

The correction terms are written as a perturbative expansion:
\begin{equation}
\Delta_{ij}(x) = \Delta_{ij}^{(0)}(x)
+ \api\,\Delta_{ij}^{(1)}(x) +
\left(\api\right)^2\,\Delta_{ij}^{(2)}(x)
+ \order{\alpha_s^3}\,.
\end{equation}
Explicit dependence on the number of active flavors $n_f$ appears only
at \nnlo.  Because the terms are large and cumbersome, it is
convenient to write
\begin{equation}
\Delta_{ij}^{(2)}(x) = \Delta_{ij}^{(2)\,A}(x)
    + n_f\,\Delta_{ij}^{(2)\,F}(x)\,.
\end{equation}
All results will be presented for the scale choices
$\muF=\muR=\mhiggs$.  The corresponding expressions for general values
of $\muF$ and $\muR$ can be reconstructed from renormalization scale
invariance of the partonic, and factorization invariance of the
hadronic cross section.\footnote{ The analytic results including all
scale dependences can also be obtained from the authors upon request.}
\vfil\eject
\begin{widetext}

\subsection{The $\bbar$ subprocess}
In the \vfs{} approach, the tree-level $\bbar$ annihilation term is the
\lo{} contribution.  Thus, this is the only term for which
$\Delta_{ij}^{(0)}(x)$ does not vanish.  The \lo{} contribution to
$\bbar\to\phi+X$ is
\begin{equation}
\Delta_{\bbar}^{(0)}(x) = \delta(1-x).
\end{equation}
The \nlo{} contribution is
\begin{equation}
\begin{split}
\Delta_{\bbar}^{(1)}(x) = &\ 
       - \frac{4 - 8\,\zeta_2}{3} \, \delta(1-x)
       + \frac{16}{3} \, \Di1(1-x)
       - \frac{16 + 8\,x + 8\,x^2}{3} \, \ln\Lx 1-x\Rx
       + \frac{4\,x - 4\,x^2}{3}
       - \frac{8}{3} \, \frac{\ln\Lx x \Rx}{1-x}
       + \frac{8 + 4\,x + 4\,x^2}{3} \, \ln\Lx x \Rx\,,
\end{split}
\end{equation}
where
$\displaystyle\Di{n}(1-x)\equiv\LB\frac{\ln^{n}(1-x)}{1-x}\RB_+$, and
$\zeta_2 \equiv \pi^2/6 \approx 1.64493$, $\zeta_3\approx 1.20206$.

At \nnlo, the contributions are ($\zeta_4 \equiv \pi^4/90 \approx
1.08232$)
\begin{equation}
\begin{split}
   \Delta_{\bbar}^{(2)\,A} =&\
       \frac{115 + 116\,\zeta_2 - 156\,\zeta_3 - 19\,\zeta_4}{18}\,
                  \delta\Lx 1-x \Rx
       - \frac{404 - 396\,\zeta_2 - 1146\,\zeta_3}{27} \, \Di0\Lx 1-x \Rx\\&
       + \frac{204 - 200\,\zeta_2}{9} \, \Di1\Lx 1-x \Rx
       - \frac{44}{3} \, \Di2\Lx 1-x \Rx
       + \frac{128}{9} \, \Di3\Lx 1-x \Rx\\&
       - \frac{128 + 64\,x + 64\,x^2}{9} \, \ln^3\Lx 1-x \Rx
       + \frac{140 + 40\,x + 92\,x^2 - 8\,x^3}{9} \, \ln^2\Lx 1-x \Rx\\&
       - \frac{248}{9} \, \frac{\ln^2\Lx 1-x \Rx\,\ln\Lx x \Rx}{1 - x}
       + \frac{248 + 168\,x + 168\,x^2}{9} \,
                   \ln^2\Lx 1-x \Rx\,\ln\Lx x \Rx\\&
       - \frac{604 + 138\,x + 423\,x^2 + 44\,x^3}{27} \, \ln\Lx 1-x \Rx
       + \frac{200 + 100\,x + 100\,x^2}{9} \, \zeta_2\,\ln\Lx 1-x \Rx\\&
       + 24 \, \frac{\ln\Lx 1-x \Rx\,\ln\Lx x \Rx}{1 - x}
       - \frac{216 + 110\,x + 160\,x^2 - 24\,x^3}{9} \,
                   \ln\Lx 1-x \Rx\,\ln\Lx x \Rx\\&
       + \frac{148}{9} \, \frac{\ln\Lx 1-x \Rx\,\ln^2\Lx x \Rx}{1 - x}
       - \frac{148 + 110\,x + 110\,x^2}{9} \, \ln\Lx 1-x \Rx\,\ln^2\Lx x \Rx\\&
       + \frac{20}{9} \, \frac{\ln\Lx 1-x \Rx\,\Li2\Lx 1-x \Rx}{1 - x}
       - \frac{20 - 78\,x - 78\,x^2}{9} \, \ln\Lx 1-x \Rx\,\Li2\Lx 1-x \Rx\\&
       + \frac{4640 + 1017\,x + 2958\,x^2 + 721\,x^3}{324}
       - \frac{140 - 15\,x + 147\,x^2 - 8\,x^3}{9} \, \zeta_2\\&
       - \frac{382 + 191\,x + 191\,x^2}{9} \, \zeta_3
       - \frac{146}{9} \, \frac{\ln\Lx x \Rx}{1 - x}
       + \frac{164}{9} \, \frac{\zeta_2\,\ln\Lx x \Rx}{1 - x}
       - \frac{23}{3} \, \frac{\ln^2\Lx x \Rx}{1 - x}
       - \frac{44}{27} \, \frac{\ln^3\Lx x \Rx}{1 - x}\\&
       + \frac{876 + 249\,x + 444\,x^2 + 38\,x^3}{54} \, \ln\Lx x \Rx
       - \frac{164 + 126\,x + 126\,x^2}{9} \, \zeta_2\,\ln\Lx x \Rx\\&
       + \frac{138 + 73\,x + 115\,x^2 - 12\,x^3}{18} \, \ln^2\Lx x \Rx
       + \frac{44 + 21\,x + 21\,x^2 + 4\,x^3}{27} \, \ln^3\Lx x \Rx\\&
       + \frac{\Li2\Lx 1-x \Rx}{1 - x}
       + \frac{58}{9} \, \frac{\Li2\Lx 1-x \Rx\,\ln\Lx x \Rx}{1 - x}
       - \frac{142}{9} \, \frac{\Li3\Lx 1-x \Rx}{1 - x}
       - \frac{64}{9\,\Lx 1-x \Rx} \, \Li3\Lx - \frac{1 - x}{x} \Rx\\&
       + \frac{7 - 51\,x - 10\,x^2 + 10\,x^3}{9} \, \Li2\Lx 1-x \Rx
       - \frac{58 + 88\,x + 88\,x^2 + 2\,x^3}{9} \,
                   \Li2\Lx 1-x \Rx\,\ln\Lx x \Rx\\&
       - \frac{x - 2\,x^2 - 2\,x^3}{9} \, \Li2\Lx 1-x^2 \Rx
       + \frac{x^3}{3} \, \Li2\Lx 1-x^2 \Rx\,\ln\Lx x \Rx\\&
       + \frac{142 + 37\,x + 37\,x^2 + 6\,x^3}{9} \, \Li3\Lx 1-x \Rx
       + \frac{64 + 94\,x + 94\,x^2 - 6\,x^3}{9} \,
                   \Li3\Lx - \frac{1 - x}{x} \Rx\\&
       - \frac{7\,x^3}{18} \, \Li3\Lx 1-x^2 \Rx
       - \frac{x^3}{18} \, \Li3\Lx - \frac{1 - x^2}{x^2} \Rx
       - \frac{2\,x^3}{3} \, \LB \Li3\Lx \frac{1 - x}{1 + x} \Rx
                   - \Li3\Lx - \frac{1 - x}{1 + x} \Rx\RB\,,
\end{split}
\end{equation}
\begin{equation}
\begin{split}
   \Delta_{\bbar}^{(2)\,F} =&\
       \frac{2  - 10\,\zeta_2 + 18\,\zeta_3}{27} \, \delta\Lx 1-x \Rx
       + \frac{56 - 72\,\zeta_2}{81} \, \Di0\Lx 1-x \Rx
       - \frac{40}{27} \, \Di1\Lx 1-x \Rx
       + \frac{8}{9} \, \Di2\Lx 1-x \Rx\\&
       - \frac{8 + 4\,x + 4\,x^2}{9} \,\ln^2\Lx 1-x \Rx
       + \frac{40 + 8\,x + 32\,x^2}{27} \,\ln\Lx 1-x \Rx
       - \frac{16}{9} \, \frac{\ln\Lx 1-x \Rx\,\ln\Lx x \Rx}{1 - x}\\&
       + \frac{16 + 8\,x + 8\,x^2}{9} \,\ln\Lx 1-x \Rx\,\ln\Lx x \Rx
       + \frac{10}{9} \, \frac{\ln\Lx x \Rx}{1 - x}
       + \frac{2}{3} \, \frac{\ln^2\Lx x \Rx}{1 - x}
       - \frac{2}{9} \, \frac{\Li2\Lx 1-x \Rx}{1 - x}
       - \frac{56 + x + 55\,x^2}{81} \\&
       + \frac{8 + 4\,x + 4\,x^2}{9} \, \zeta_2
       - \frac{10 + 3\,x + 7\,x^2}{9} \,\ln\Lx x \Rx
       - \frac{12 + 7\,x + 7\,x^2}{18} \,\ln^2\Lx x \Rx
       + \frac{2}{9} \, \Li2\Lx 1-x \Rx\,.
\end{split}
\end{equation}
Note that the $\Di{n}$ terms in this result could also be derived by
other methods~\cite{Kramer:1998iq,Vogt:2000ci,Harlander:2001is,
Catani:2001ic,Kidonakis:2003tx,Magnea:1991qg,Contopanagos:1997nh}.

\subsection{The $bg$ subprocess}
The $bg\to\phi+X$ subprocess first enters at \nlo, where the
contribution is
\begin{equation}
\begin{split}
   &\Delta_{bg}^{(1)} = \Delta_{\bar{b}g}^{(1)} =
       \frac{x - 2\,x^2 + 2\,x^3}{2} \, \ln\Lx 1-x\Rx
       - \frac{3\,x - 10\,x^2 + 7\,x^3}{8}
       - \frac{x - 2\,x^2 + 2\,x^3}{4} \, \ln\Lx x \Rx\,.
\end{split}
\end{equation}

At \nnlo, the contribution is
\begin{equation}
\begin{split}
   &\Delta_{bg}^{(2)\,A} = \Delta_{\bar{b}g}^{(2)\,A} =\\&\phantom{+}
       \frac{257\,x - 514\,x^2 + 514\,x^3}{144} \, \ln^3\Lx 1-x \Rx\\&
       + \frac{16 - 59\,x + 272\,x^2 - 237\,x^3}{16} \, \ln^2\Lx 1-x \Rx
       - \frac{11\,x - 94\,x^2 + 62\,x^3}{8} \,
                   \ln^2\Lx 1-x \Rx\,\ln\Lx x \Rx\\&
       + \frac{16 + 28\,x - 731\,x^2 + 726\,x^3}{48} \, \ln\Lx 1-x \Rx
       - \frac{35\,x - 70\,x^2 + 70\,x^3}{12} \, \zeta_2\,\ln\Lx 1-x \Rx\\&
       + \frac{65\,x - 508\,x^2 + 774\,x^3}{24} \,
                   \ln\Lx 1-x \Rx\,\ln\Lx x \Rx
       - \frac{3\,x + 174\,x^2 - 98\,x^3}{24} \,
                   \ln\Lx 1-x \Rx\,\ln^2\Lx x \Rx\\&
       + \frac{77\,x + 134\,x^2 - 86\,x^3}{24} \,
                   \ln\Lx 1-x \Rx\,\Li2\Lx 1-x \Rx
       + \frac{3\,x + 6\,x^2 + 6\,x^3}{4} \,
                   \ln\Lx 1-x \Rx\,\Li2\Lx 1-x^2 \Rx\\&
       - \frac{208 - 411\,x - 1350\,x^2 + 1781\,x^3}{288}
       - \frac{16 - 31\,x + 176\,x^2 - 169\,x^3}{16} \, \zeta_2
       + \frac{161\,x - 322\,x^2 + 322\,x^3}{48} \, \zeta_3\\&
       + \frac{32\,x + 536\,x^2 - 993\,x^3}{48} \, \ln\Lx x \Rx
       + \frac{7\,x - 230\,x^2 + 134\,x^3}{24} \, \zeta_2\,\ln\Lx x \Rx\\&
       - \frac{47\,x - 604\,x^2 + 1028\,x^3}{96} \, \ln^2\Lx x \Rx
       - \frac{35\,x + 146\,x^2 + 12\,x^3}{144} \, \ln^3\Lx x \Rx\\&
       + \frac{48 - 43\,x + 152\,x^2 + 236\,x^3}{24} \, \Li2\Lx 1-x \Rx
       - \frac{10\,x + 34\,x^2 - 17\,x^3}{6} \,
                   \Li2\Lx 1-x \Rx\,\ln\Lx x \Rx\\&
       - \frac{19\,x + 40\,x^2 + 21\,x^3}{48} \, \Li2\Lx 1-x^2 \Rx
       - \frac{9\,x + 18\,x^2 + 10\,x^3}{12} \,
                   \Li2\Lx 1-x^2 \Rx\,\ln\Lx x \Rx\\&
       - \frac{14\,x + 8\,x^2 + 9\,x^3}{3} \, \Li3\Lx 1-x \Rx
       + \frac{13\,x + 118\,x^2 - 18\,x^3}{12} \,
                   \Li3\Lx - \frac{1 - x}{x} \Rx\\&
       + \frac{9\,x + 18\,x^2 - 14\,x^3}{48} \, \Li3\Lx 1-x^2 \Rx
       + \frac{9\,x + 18\,x^2 - 14\,x^3}{48} \,
                   \Li3\Lx - \frac{1 - x^2}{x^2} \Rx\\&
       + \frac{3\,x + 6\,x^2 + 6\,x^3}{2} \,
                   \LB \Li3\Lx \frac{1 - x}{1 + x} \Rx
                   - \Li3\Lx - \frac{1 - x}{1 + x} \Rx\RB\,,
\\[1em]
\Delta_{bg}^{(2)\,F} = &\ \Delta_{\bar bg}^{(2)\,F} = 0\,.
\end{split}
\end{equation}

\subsection{The $bq$ subprocess}
All remaining components contribute only at \nnlo{} and beyond.  The
contribution to $bq\to\phi+X$, where $q$ is a light ($u$,$d$,$s$ or
$c$) quark, is:
\begin{equation}
\begin{split}
   \Delta_{bq}^{(2)\,A} = &\ \Delta_{\bar{b}q}^{(2)\,A} = 
   \Delta_{b\bar{q}}^{(2)\,A} = \Delta_{\bar{b}\bar{q}}^{(2)\,A} =
       \frac{4 + 3\,x - 3\,x^2 - 4\,x^3}{9} \, \ln^2\Lx 1-x \Rx\\&
       + \frac{2\,x + 2\,x^2}{3} \, \ln^2\Lx 1-x \Rx\,\ln\Lx x \Rx
       + \frac{4 - 57\,x + 75\,x^2 - 22\,x^3}{27} \, \ln\Lx 1-x \Rx\\&
       - \frac{x - 4\,x^2 - 4\,x^3}{3} \, \ln\Lx 1-x \Rx\,\ln\Lx x \Rx
       - \frac{2\,x + 2\,x^2}{3} \, \ln\Lx 1-x \Rx\,\ln^2\Lx x \Rx\\&
       + \frac{4\,x + 4\,x^2}{3} \, \ln\Lx 1-x \Rx\,\Li2\Lx 1-x \Rx
       - \frac{208 - 915\,x + 1410\,x^2 - 703\,x^3}{648}\\&
       - \frac{4 + 3\,x - 3\,x^2 - 4\,x^3}{9} \, \zeta_2
       + \frac{93\,x - 264\,x^2 + 20\,x^3}{108} \, \ln\Lx x \Rx
       - \frac{2\,x + 2\,x^2}{3} \, \zeta_2\,\ln\Lx x \Rx\\&
       - \frac{3\,x + 15\,x^2 + 40\,x^3}{72} \, \ln^2\Lx x \Rx
       + \frac{x + x^2}{36} \, \ln^3\Lx x \Rx
       + \frac{16 - 3\,x + 21\,x^2 + 8\,x^3}{18} \, \Li2\Lx 1-x \Rx\\&
       - \Lx x + x^2 \Rx \, \Li2\Lx 1-x \Rx\,\ln\Lx x \Rx
       - \frac{2\,x + 2\,x^2}{3} \, \Li3\Lx 1-x \Rx
       + \frac{2\,x + 2\,x^2}{3} \, \Li3\Lx - \frac{1 - x}{x} \Rx\,,
\\[1em]
\Delta_{bq}^{(2)\,F} = &\ \Delta_{\bar{b}q}^{(2)\,F} = 
\Delta_{b\bar q}^{(2)\,F} = \Delta_{\bar{b}\bar q}^{(2)\,F} = 0\,.
\end{split}
\end{equation}

\subsection{The $gg$ subprocess}
The contribution to $gg\to\phi+X$ is:
\begin{equation}
\begin{split}
  \Delta_{gg}^{(2)\,A} =&\
       - \Lx x + 2\,x^2 - 3\,x^3 \Rx \, \ln^2\Lx 1-x \Rx
       - \frac{x + 4\,x^2 + 4\,x^3}{2} \, \ln^2\Lx 1-x \Rx\,\ln\Lx x \Rx\\&
       + \frac{23\,x + 52\,x^2 - 75\,x^3}{8} \, \ln\Lx 1-x \Rx
       + \frac{5\,x + 16\,x^2 - 4\,x^3}{4} \, \ln\Lx 1-x \Rx\,\ln\Lx x \Rx\\&
       + \frac{x + 4\,x^2 + 4\,x^3}{4} \, \ln\Lx 1-x \Rx\,\ln^2\Lx x \Rx
       - \Lx x + 4\,x^2 + 4\,x^3 \Rx \, \ln\Lx 1-x \Rx\,\Li2\Lx 1-x \Rx\\&
       - \frac{163\,x + 1528\,x^2 - 1691\,x^3}{128}
       + \Lx x + 2\,x^2 - 3\,x^3 \Rx \, \zeta_2\\&
       - \frac{54\,x + 312\,x^2 - 223\,x^3}{64} \, \ln\Lx x \Rx
       + \frac{x + 4\,x^2 + 4\,x^3}{2} \, \zeta_2\,\ln\Lx x \Rx\\&
       - \frac{16\,x + 111\,x^2 - 43\,x^3}{64} \, \ln^2\Lx x \Rx
       + \frac{7\,x + 25\,x^2 + 34\,x^3}{48} \, \ln^3\Lx x \Rx\\&
       - \frac{4\,x - 15\,x^2 - 62\,x^3}{16} \, \Li2\Lx 1-x \Rx
       + \frac{11\,x + 44\,x^2 + 30\,x^3}{16} \,
                   \Li2\Lx 1-x \Rx\,\ln\Lx x \Rx\\&
       + \frac{x^2 - 6\,x^3}{32} \, \Li2\Lx 1-x^2 \Rx
       + \frac{3\,x + 6\,x^2 + 38\,x^3}{64} \,
                   \Li2\Lx 1-x^2 \Rx\,\ln\Lx x \Rx\\&
       + \frac{x + 3\,x^2 + 18\,x^3}{8} \, \Li3\Lx 1-x \Rx
       - \frac{15\,x + 60\,x^2 + 30\,x^3}{16} \,
                   \Li3\Lx - \frac{1 - x}{x} \Rx\\&
       - \frac{5\,x + 10\,x^2 + 74\,x^3}{128} \, \Li3\Lx 1-x^2 \Rx
       - \frac{3\,x + 6\,x^2 + 70\,x^3}{128} \,
                   \Li3\Lx - \frac{1 - x^2}{x^2} \Rx\\&
       - \frac{x + 2\,x^2 + 2\,x^3}{32} \,
                   \LB \Li3\Lx \frac{1 - x}{1 + x} \Rx
                    - \Li3\Lx - \frac{1 - x}{1 + x} \Rx\RB\,,
\\[1em]
  \Delta_{gg}^{(2)\,F} =&\ 0\,.
\end{split}
\end{equation}

\subsection{The $bb$ subprocess}
The contribution to $bb\to\phi+X$ is:
\begin{equation}
\begin{split}
   \Delta_{bb}^{(2)\,A} =&\ \Delta_{\bar{b}\bar{b}}^{(2)\,A} =
        \frac{8 + 6\,x - 6\,x^2 - 8\,x^3}{9} \, \ln^2\Lx 1-x \Rx
       + \frac{4\,x + 4\,x^2}{3} \, \ln^2\Lx 1-x \Rx\,\ln\Lx x \Rx\\&
       + \frac{8 - 138\,x + 174\,x^2 - 44\,x^3}{27} \, \ln\Lx 1-x \Rx
       - \frac{10\,x - 20\,x^2 - 24\,x^3}{9} \, \ln\Lx 1-x \Rx\,\ln\Lx x \Rx\\&
       + \frac{4}{9} \, \frac{\ln\Lx 1-x \Rx\,\ln^2\Lx x \Rx}{1 + x}
       - \frac{16}{9} \, \frac{\ln\Lx 1-x \Rx\,\Li2\Lx 1-x \Rx}{1 + x}
       + \frac{8}{9} \, \frac{\ln\Lx 1-x \Rx\,\Li2\Lx 1-x^2 \Rx}{1 + x}\\&
       - \frac{4 + 10\,x + 14\,x^2}{9} \, \ln\Lx 1-x \Rx\,\ln^2\Lx x \Rx
       + \frac{16 + 16\,x + 32\,x^2}{9} \, \ln\Lx 1-x \Rx\,\Li2\Lx 1-x \Rx\\&
       - \frac{8 - 4\,x + 4\,x^2}{9} \, \ln\Lx 1-x \Rx\,\Li2\Lx 1-x^2 \Rx
       - \frac{52 - 357\,x + 510\,x^2 - 205\,x^3}{81}\\&
       - \frac{8 + 6\,x - 6\,x^2 - 8\,x^3}{9} \, \zeta_2
       + \frac{117\,x - 279\,x^2 + 20\,x^3}{54} \, \ln\Lx x \Rx
       - \frac{4\,x + 4\,x^2}{3} \, \zeta_2\,\ln(x)\\&
       - \frac{x + 11\,x^2 + 34\,x^3}{36} \, \ln^2(x)
       + \frac{28 - 12\,x + 17\,x^2}{54} \, \ln^3\Lx x \Rx
       - \frac{14}{27} \, \frac{\ln^3\Lx x \Rx}{1 + x}\\&
       + \frac{4}{3} \, \frac{\Li2\Lx 1-x \Rx\,\ln\Lx x \Rx}{1 + x}
       - \frac{8}{9} \, \frac{\Li2\Lx 1-x^2 \Rx\,\ln\Lx x \Rx}{1 + x}
       - \frac{4}{3} \, \frac{\Li3\Lx 1-x \Rx}{1 + x}
       + \frac{4}{3\,\Lx 1+x \Rx} \, \Li3\Lx - \frac{1 - x}{x} \Rx\\&
       + \frac{1}{3} \, \frac{\Li3\Lx 1-x^2 \Rx}{1 + x}
       + \frac{1}{9\,\Lx 1+x \Rx} \, \Li3\Lx - \frac{1 - x^2}{x^2} \Rx
       + \frac{20}{9\,\Lx 1+x \Rx} \, \LB \Li3\Lx \frac{1 - x}{1 + x} \Rx
                   - \Li3\Lx - \frac{1 - x}{1 + x} \Rx\RB\\&
       + \frac{16 - 12\,x + 16\,x^2 + 11\,x^3}{9} \, \Li2\Lx 1-x \Rx
       - \frac{12 + 8\,x + 29\,x^2}{9} \, \Li2\Lx 1-x \Rx\,\ln\Lx x \Rx\\&
       + \frac{x}{9} \, \Li2\Lx 1-x^2 \Rx
       + \frac{16 - 11\,x + 11\,x^2}{18} \, \Li2\Lx 1-x^2 \Rx\,\ln\Lx x \Rx
       + \frac{12 - 27\,x + 5\,x^2}{9} \, \Li3\Lx 1-x \Rx\\&
       - \frac{12 - 16\,x - 9\,x^2 - 2\,x^3}{9} \,
                   \Li3\Lx - \frac{1 - x}{x} \Rx
       - \frac{12 - 13\,x + 13\,x^2}{36} \, \Li3\Lx 1-x^2 \Rx\\&
       - \frac{4 - 3\,x + 3\,x^2}{36} \, \Li3\Lx - \frac{1 - x^2}{x^2} \Rx
       - \frac{20 - 13\,x + 13\,x^2}{9} \,
                   \LB \Li3\Lx \frac{1 - x}{1 + x} \Rx - \Li3\Lx
                   - \frac{1 - x}{1 + x} \Rx\RB\,,
\\[1em]
\Delta_{bb}^{(2)\,F} =&\ \Delta_{\bar{b}\bar{b}}^{(2)\,F} = 0\,.
\end{split}
\end{equation}

\subsection{The $\qqbar$ subprocess}
The contribution to $\qqbar\to\phi+X$ is ($q\in\{u,d,s,c\}$):
\begin{equation}
\begin{split}
   \Delta_{q\bar{q}}^{(2)\,A} = &\ 
       - \frac{2\,x - 8\,x^2 + 6\,x^3}{9}
       - \frac{x - 2\,x^2 - 3\,x^3}{9} \, \ln\Lx x \Rx
       + \frac{x^3}{9} \, \ln^2\Lx x \Rx\hskip 1in\\&\ 
       - \frac{4\,x^3}{9} \, \Li2\Lx 1-x \Rx
       + \frac{2\,x^3}{9} \, \Li2\Lx 1-x^2 \Rx\,,\\[1em]
   \Delta_{q\bar{q}}^{(2)\,F} = &\ 0\,.
\end{split}
\end{equation}

\vskip 2cm{}
\end{widetext}
\end{appendix}


\providecommand{\href}[2]{#2}\begingroup\raggedright\endgroup


\end{document}